\documentclass[%
 reprint,
 amsmath,amssymb,
 aps,
]{revtex4-2}

\usepackage{xcolor}
\usepackage{graphicx}
\usepackage{dcolumn}
\usepackage{bm}

\usepackage[caption=false]{subfig}

\graphicspath{{./}{figures/}}

\begin{document}

\title{Evolution of coupled weakly-driven waves in a dissipative plasma}

\author{N. M. Pham}
\email{npham@pppl.gov}
\affiliation{Princeton Plasma Physics Laboratory, Princeton University, Princeton, NJ, 08543}

\author{V. N. Duarte}
\email{vduarte@pppl.gov}
\affiliation{Princeton Plasma Physics Laboratory, Princeton University, Princeton, NJ, 08543}

\begin{abstract}
    The nonlinear collisional dynamics of coupled driven plasma waves in the presence of background dissipation is studied analytically within kinetic theory. Sufficiently near marginal stability, phase space correlations are poorly preserved and time delays become unimportant. The system is then shown to be governed by two first-order coupled autonomous differential equations of cubic order for the wave amplitudes and two complementary first-order equations for the evolution of their phases. That system of equations can be decoupled and further simplified to a single second-order differential equation of Liénard's type for each amplitude. Numerical solutions for this equation are obtained in the general case while analytic solutions are obtained for special cases in terms of parameters related to the spacing of the resonances of the two waves in frequency space, e.g., wave lengths and oscillation frequencies. These parameters are further analyzed to find classes of  quasi-steady saturation and pulsating scenarios. To classify equilibrium points, local stability analysis is applied, and bifurcation conditions are determined. When the two waves saturate at similar amplitude levels, their  combined signal is shown to invariably  exhibit amplitude beating and phase jumps of nearly $\pi$. The obtained analytical results can be used to benchmark simulations and to interpret eigenmode amplitude measurements in fusion experiments.
\end{abstract}

\maketitle

\section{Introduction}

In most kinetic plasma systems of practical interest, wave dynamics is intrinsically of coupled nature. For instance, in studies of the drift wave turbulence \cite{kadomtsev1965plasma,diamond2010modern}, quantum two-stream instability \cite{haasquantum}, and Alfvénic eigenmodes in fusion plasmas \cite{mikhailovskii2017instabilities}, the coupling between modes plays a central role in the wave amplitude evolution and the underlying particle transport. Although the resonant dynamics of a single plasma wave has been relatively well understood in analytic terms \cite{ONiel1965,Mazitov1965,BerkBreizman1990a}, the dynamics of coupled modes has proven to be more intricate \cite{sagdeev1969nonlinear,stix1992waves,sagdeev1988nonlinear}.

The purpose of this work is to analytically study the basic nonlinear dynamics of two marginally-unstable overlapping plasma waves, using the framework of Ref. \cite{Zalesny,BerkPRL1996} as a starting point (extension studies have been also reported in Refs. \cite{Galant_2011,Galant+2013+1598+1604}). We show that one can achieve considerable analytic simplification for experimentally-relevant scenarios  when phase memory of resonant particles is poorly retained. In this case, the dynamics is shown to be ultimately governed by a system of coupled Landau-Stuart equations. We characterize the main properties of such a system, in terms of saturation levels, stability fixed points, beating and phase jumps, and compare the analytic results modeling poor phase memory with simulations for the case in which the full phase information is retained.

We consider the case in which the two oscillators interact with each other indirectly, i.e., via their common medium. This occurs in a wide range of applications of the nonlinear dynamics of oscillators (see Ref. \cite{Sharma2012} and Refs. 24-31 therein). In plasmas, this setup is applicable within the realm of weak turbulence theory with a wave-particle type of nonlinearity \cite{sagdeev1969nonlinear} between the modes and the distribution function of the resonating minority species. Explicitly, this coupling exists entirely through particles that resonate with both modes. Although restricted to weak nonlinearities where the modes are not magnetodynamically (wave-wave) coupled, our treatment offers new and transparent analytic intuition and insights on the evolution of coupled driven modes of discrete nature. Our results are directly relevant to extending resonance-broadened quasilinear models \cite{duarte2019collisional} and multi-mode Alfvén eigenmode simulations in plasmas \cite{Schneller_2016,FitzgeraldITER_NF2016} and overlap-mediated redistribution in galactic dynamics \cite{tagger1987nonlinear,sygnet,masset1999non,hamilton2022galactic,hamilton2023evolution}, and are directly useful to simulations of the transport induced by plasma eigenmodes \cite{GorelenkovDuarteNF2018}. 

This paper is structured as follows. In Sec. II, the governing nonlinear, time-delayed integro-differential equations are presented. When resonant particles decorrelate from a resonance on a timescale shorter than the characteristic wave growth time, those equations are shown to reduce to a set of time-local ordinary differential equations, or equivalently, to a single Liénard's equation, which can be explicitly solved in several regimes. In Sec. III, saturation levels and stability fixed-point diagrams are categorized. Sec. IV shows that the analyzed system leads to wave beating and phase jumps when the two amplitudes are similar. A synchronicity condition is also determined. Sec. V shows numerical results for the coupled wave evolution. It presents comparisons between the reduced time-local system against the more complex time-delayed system for a range of collisionalities.  Discussions on the implications of our findings are presented in Sec. VI. Details of the calculations and numerical scheme utilized are shown in the Appendix.

\section{Theoretical framework} 

\subsection{Model equations}

We start with the evolution equations for mode amplitudes $\hat{A_j}$ $(j=1,2)$ derived by \cite{Zalesny} for the case in which each mode is primarily driven by one dominant resonance. For a Krook collision operator of the form $(df/dt)_{\text{coll}} = \nu (F_{0}-f)$ where $\nu$ is an effective collision frequency and $F_{0}$ is the equilibrium distribution function, in the absence of perturbation \footnote{The equations presented in this paper are not exactly the same as the equations found in \cite{Zalesny}. Instead, they adhere to the original normalization of Berk \textit{et al} \cite{BerkPRL1996}. The conversion between Berk's and Zalesny's normalization is $2 A_{\text{Zalesny}} = A_{\text{Berk}}$.}, the evolution equations for near-threshold modes are


\begin{widetext}
\begin{equation} \label{amp1}
\begin{aligned}
\frac{\mathrm{d} \hat{A_1}}{\mathrm{d} t}= & \hat{A_1} - \frac{1}{2} \int_0^{t/2} \mathrm{d}\eta\int_0^{t - 2\eta}\mathrm{d}\chi \cdot  \Bigl[\eta^2 \cdot \bigl(\hat{A_1}(t - \eta)\hat{A_1}(t - \eta - \chi)\hat{A_1^*}(t - 2\eta - \chi)\\
& + \hat{A_1}(t- \eta)\hat{A_2}(t - \eta - \chi)\hat{A_2^*}(t - 2 \eta - \chi) \cdot e^{-ip_1\eta}\bigr) \\
& + \hat{A_2}(t - \eta)\hat{A_1}(t - \eta - \chi)\hat{A_2^*}(t - 2\eta - \chi) e^{-ip_1(2\eta + \chi)} \cdot \eta (\eta + u_1(\eta + \chi ))\Bigr] \cdot e^{-\hat{\nu} (2\eta + \chi)}
\end{aligned}
\end{equation}
and
\begin{equation} \label{amp2}
\begin{aligned}
\frac{\mathrm{d} \hat{A_2}}{\mathrm{d} t}= & \hat{A_2} - \frac{1}{2} \int_0^{t/2} \mathrm{d}\eta\int_0^{t - 2\eta}\mathrm{d}\chi \cdot  \bigl[\eta^2 \cdot \bigl(\hat{A_2}(t - \eta)\hat{A_2}(t - \eta - \chi)\hat{A_2^*}(t - 2\eta - \chi)\\
& + \hat{A_2}(t- \eta)\hat{A_1}(t - \eta - \chi)\hat{A_1^*}(t - 2 \eta - \chi) \cdot e^{-ip_2\eta}\bigr) \\
& + \hat{A_1}(t - \eta)\hat{A_2}(t - \eta - \chi)\hat{A_1^*}(t - 2\eta - \chi) e^{-ip_2(2\eta + \chi)} \cdot \eta (\eta + u_2(\eta + \chi ))\bigr] \cdot e^{-\hat{\nu} (2\eta + \chi)},
\end{aligned}
\end{equation}
\end{widetext}
where $u_j \equiv \frac{\Delta k}{k_j}$, $p_j \equiv  \frac{\omega_j}{\gamma}\left(\frac{\Delta k}{k_j} - \frac{\Delta \omega}{\omega_j}\right)$, $\Delta k \equiv k_1 - k_2$, and $\Delta \omega  \equiv \omega_1 - \omega_2$.  $k_j$ and $\omega_j$ represent the wave vector amplitudes and the oscillation frequency of each wave respectively.

It is convenient to normalize the effective collisional frequency with the wave net growth rate, $\hat{\nu}=\nu/(\gamma_L-\gamma_d)$, where $\gamma_L$ is the linear kinetic driving rate due to the resonating minority species and $\gamma_d$ is the dissipation (damping) rate from the background plasma. Time is also normalized with the wave inverse net growth rate $(\gamma_L-\gamma_d)^{-1}$. Eqs. ~\ref{amp1} and \ref{amp2} are constructed under the assumption that $|\Delta k/k_j|\ll1$ and $|\Delta \omega/\omega_j|\ll1$ \cite{Zalesny}. For these equations, the waves are assumed sufficiently near marginal stability which requires that  $\gamma \equiv \gamma_L - \gamma_d \ll \gamma_L$. Physically, this regime is characterized by strong energy conversions and are shown to be experimentally applicable, for instance, in systems with plasma self-heating or $\alpha$-energy channeling. In our case, the free energy injected into the system is readily converted to thermal energy via strong collisions which, in turn, heat the background plasma.

\subsection{Stochasticity-dominated limit}
In order to make analytical progress, in this work we consider the limit in which stochasticity dominates the dynamics \cite{duarte2018analytical}, i.e., when collisions regulate the wave growth. In this experimentally relevant scenario \cite{DuarteAxivPRL,duarte2020shifting} in which the effective collision frequency is much greater than the net rate of growth ($\hat{\nu}=\nu/(\gamma_L-\gamma_d)\gg1$), resonant particles receive frequent random kicks and have poor phase memory retention. In this case, the exponential kernel of Eqs. \ref{amp1} and \ref{amp2} forces the integrand to be virtually zero except where both $\eta$ and $\chi$ are close to zero. This limit leads to considerable simplification. Firstly, the arguments inside $\hat{A_j}$ within the integrand can be evaluated time locally around the peak location of the integrand and can be extracted from the integral \cite{duarte2018analytical}. Secondly, the integration bounds can be extended to infinity when the following change of variable $x = \hat{\nu}\eta$ is applied. Lastly, the resulting expressions can be explicitly integrated (for details, refer to the derivation shown in Appendix A). This implies that Eqs. \ref{amp1} and \ref{amp2} simplify into the following pair of coupled differential equations,

 
\begin{equation} \label{eq:comampeq}
    \begin{aligned}
        & \frac{\mathrm{d} \hat{A_1}}{\mathrm{d} t} = \hat{A_1} \left( 1 - b_0|\hat{A_1}|^2 - b_1|\hat{A_2}|^2\right), \\
        & \frac{\mathrm{d} \hat{A_2}}{\mathrm{d} t} = \hat{A_2} \left( 1 - b_0|\hat{A_2}|^2 - b_2|\hat{A_1}|^2\right). \\
    \end{aligned}
\end{equation}
The coefficients that appear in Eq. \ref{eq:comampeq} are given by 

\begin{figure}[bh!] 
  \centering
    \includegraphics[width=\linewidth]{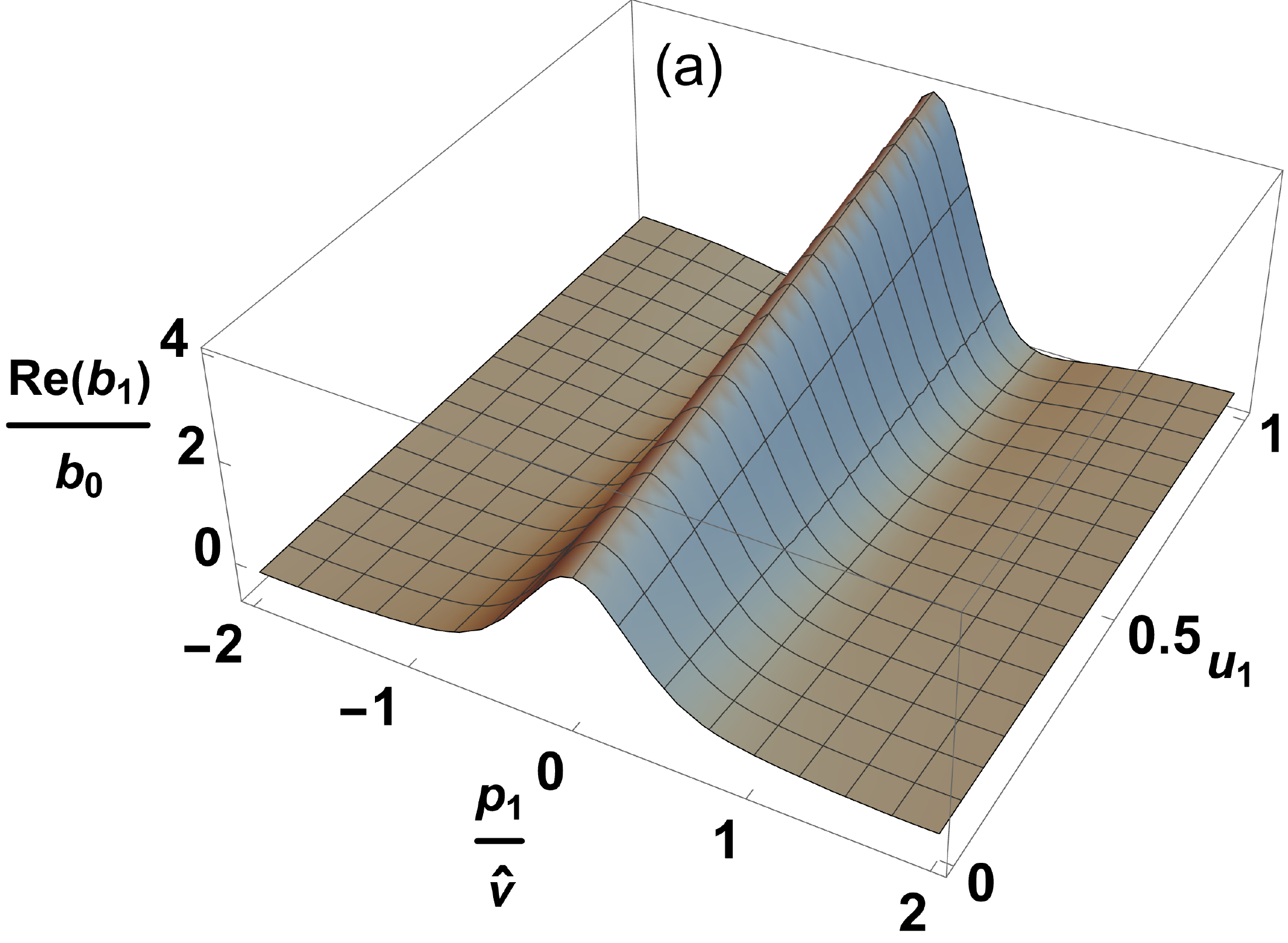}

    \includegraphics[width=\linewidth]{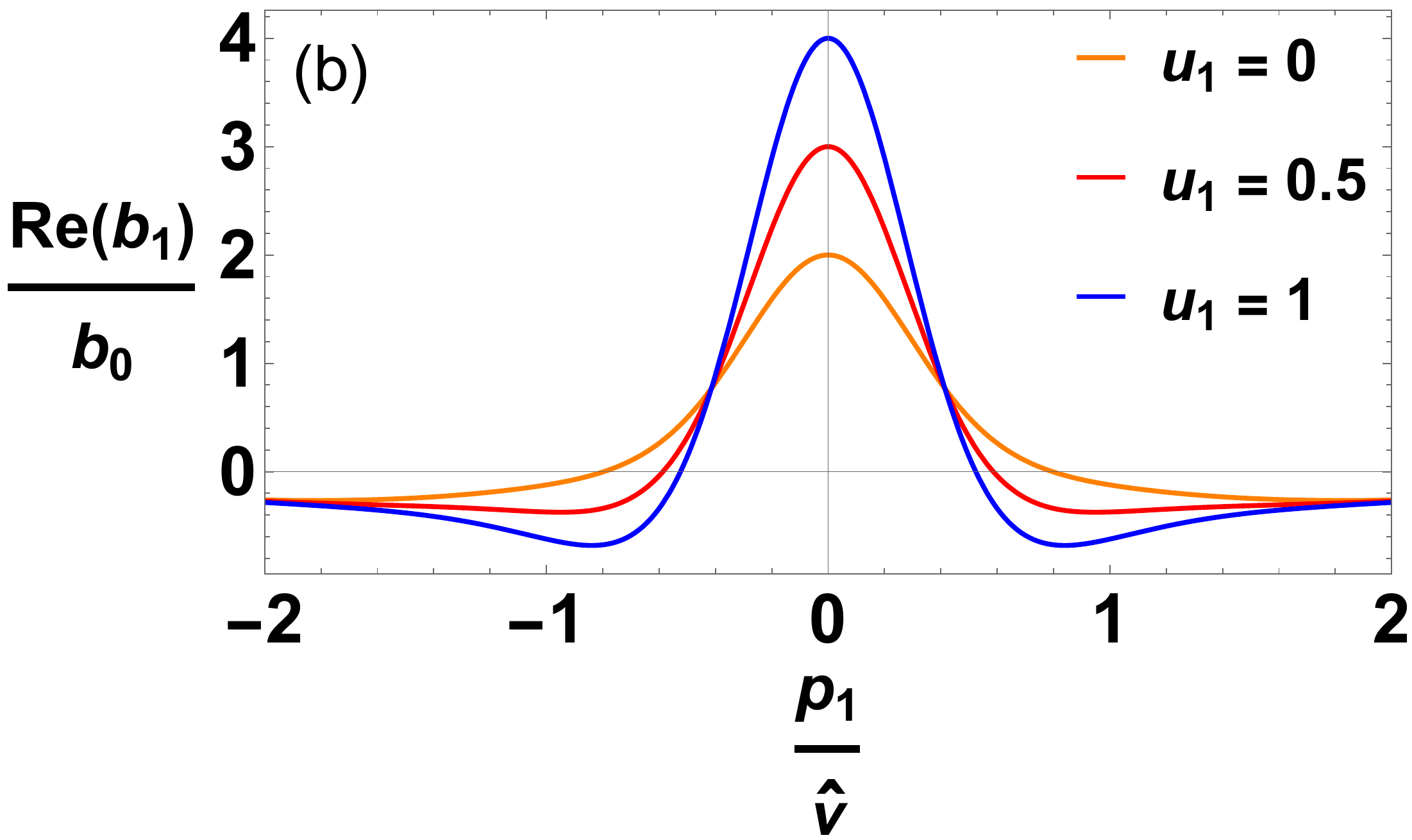}
  \caption{(a) Range of $\mathrm{Re}(b_1)/b_0$ as a function of $u_1$ and $p_1/\hat{\nu}$. As a reminder, in order for Eqs. \ref{amp1} and \ref{amp2} to be strictly valid, $|u_1|\ll1$.  (b) Cross sections of the plot shown in Fig. 1(a). As the value of $u_1$ increases, the range of $\mathrm{Re}(b_1)/b_0$ also increases.}
  \label{fig:rangeplot}
\end{figure}

\begin{widetext}
\begin{equation} \label{constants}
    \begin{aligned}
         & \mathrm{Re}(b_1) = \frac{8b_0} {\left(4+ \frac{p_1^2}{\hat{\nu}^2}\right)^{3}}\cdot\left[8 -6 \left(\frac{p_1}{\hat{\nu}}\right)^{2}\right]+ \frac{\left(1+2u_1\right) b_0}{\left(1 + \frac{p_1^2}{\hat{\nu}^2}\right)^{4}}\cdot \left[1 -6\left(\frac{p_1}{\hat{\nu}}\right)^{2} + \left(\frac{p_1}{\hat{\nu}}\right)^{4}\right], \\
         & \mathrm{Im}(b_1) = \frac{8b_0} {\left(4+ \frac{p_1^2}{\hat{\nu}^2}\right)^{3}}\cdot\left[-12 \frac{p_1}{\hat{\nu}} + \left(\frac{p_1}{\hat{\nu}}\right)^{3} \right]+ 4\frac{\left(1+2u_1\right)b_0}{\left(1 + \frac{p_1^2}{\hat{\nu}^2}\right)^{4}}\cdot \left[ -\frac{p_1}{\hat{\nu}} + \left(\frac{p_1}{\hat{\nu}}\right)^{3}\right], \\ \\
         & \mathrm{Re}(b_2) = \frac{8b_0} {\left(4+ \frac{p_2^2}{\hat{\nu}^2}\right)^{3}}\cdot\left[8 -6 \left(\frac{p_2}{\hat{\nu}}\right)^{2}\right]+ \frac{\left(1+2u_2\right)b_0}{\left(1 + \frac{p_2^2}{\hat{\nu}^2}\right)^{4}}\cdot \left[1 -6\left(\frac{p_2}{\hat{\nu}}\right)^{2} + \left(\frac{p_2}{\hat{\nu}}\right)^{4}\right], \\
         & \mathrm{Im}(b_2) = \frac{8b_0} {\left(4+ \frac{p_2^2}{\hat{\nu}^2}\right)^{3}}\cdot\left[-12 \frac{p_2}{\hat{\nu}} + \left(\frac{p_2}{\hat{\nu}}\right)^{3} \right]+ 4\frac{\left(1+2u_2\right)b_0}{\left(1 + \frac{p_2^2}{\hat{\nu}^2}\right)^{4}}\cdot \left[ -\frac{p_2}{\hat{\nu}} + \left(\frac{p_2}{\hat{\nu}}\right)^{3}\right], \\ \\
    \end{aligned}
\end{equation}

\end{widetext}
where $b_0 \equiv \left(8\hat{\nu}^4\right)^{-1}$. Additionally, the ratio $b_{j}/b_0$ is purely a function of $u_{j}$ and $p_{j}/\hat{\nu}$. The range of $\mathrm{Re}(b_1)/b_0$ is shown in Fig.~\ref{fig:rangeplot} \textemdash  ~$\mathrm{Re}(b_2)$ has an identical plot in terms of $u_{2}$ and $p_{2}/\hat{\nu}$. In the limit in which $|p_j/\hat\nu| \gg 1$ for both waves, $b_j \rightarrow 0$ and $\nu/\omega_j \ll |\Delta k/k_j - \Delta \omega/\omega_j|$ suggesting that this system is constrained to have a very small effective collision frequency and an even smaller net growth rate. Formally, the ordering for the parameters is the following: $\gamma \ll \nu \ll \omega_j|\Delta k/k_j - \Delta \omega/\omega_j| $. Under these conditions, the two modes evolve independently from each other as the interaction terms, $b_1$ and $b_2$, vanish.

In the special case where $p_j = u_j = 0$, the two modes have virtually the same wavenumber and frequency, in which case Eq.~\ref{constants} leads to $b_1 = b_2 = 2b_0$. However, having $p_j = u_j = 0$ does not reduce the system to a single mode case. Rather, the derivation for Eqs. \ref{amp1} and \ref{amp2} excludes two additional mode coupling terms and as a result, fails to represent $p_j = u_j = 0$ as a single mode \cite{Zalesny}. We see that given two waves with the same frequency and wavenumber, both the one wave and the two wave saturation pictures can be applied for analysis. To achieve a power balance equation, the wave energy for two waves is calculated using 
\begin{equation}
\begin{aligned}
    WE = \ &\frac{1}{8\pi}\int\left|\hat{A_1} + \hat{A_2}\right|^2\mathrm{d}x \\
    = \ &\frac{1}{8\pi}\int\left[|\hat{A_1}|^2 + 2\mathrm{Re}(\hat{A_1}\cdot \hat{A_2^*}) + |\hat{A_2}|^2\right]\mathrm{d}x \\
\end{aligned}
\end{equation}
where the integration parameter spans over a periodic interval for both waves. If the waves have different wavenumbers, then the integral of the cross term vanishes via an orthogonality condition. However, when the waves have the same wavenumber, such as the case for $p_j=u_j=0$, then the orthogonality argument dissolves, and the cross term now supplies a nonzero contribution to the total wave energy. This special case should instead be considered as two waves whose separations $\Delta k$ and $\Delta\omega$ are virtually zero, but still represent distinct eigenmodes \cite{Zalesny}.

\subsection{Evolution of the modes amplitudes and phases}

Substituting $\hat{A_j}(t) = |\hat{A_j}|(t) e^{i\phi_j(t)}$ in Eqs. \ref{eq:comampeq} allows the two wave amplitude equations to split into four differential equations which govern the evolution of their magnitudes and phases,

\begin{equation} \label{eq:ampeqs}
    \begin{aligned}
        & \frac{\mathrm{d}|\hat{A_1}|}{\mathrm{d}t} = |\hat{A_1}|\left(1 - b_0|\hat{A_1}|^2 - \mathrm{Re}(b_1)|\hat{A_2}|^2\right) \\
        &\ \frac{\mathrm{d}\phi_1}{\mathrm{d}t} = - \mathrm{Im}(b_1)|\hat{A_2}|^2 \\
        & \frac{\mathrm{d}|\hat{A_2|}}{\mathrm{d}t} = |\hat{A_2}|\left(1 - b_0|\hat{A_2}|^2 - \mathrm{Re}(b_2)|\hat{A_1}|^2\right) \\
        & \ \frac{\mathrm{d}\phi_2}{\mathrm{d}t} = - \mathrm{Im}(b_2)|\hat{A_1}|^2 \\
    \end{aligned}
\end{equation}

The amplitude evolution in Eq.~\ref{eq:ampeqs} has close similarities to the coupled evolution equations for vortex shedding modes in fluids \cite{Barkley,Sheard}. Interestingly, in Eq.~\ref{eq:ampeqs}, the amplitude of one mode is found to regulate the phase of another. The structure of the phase evolution in Eq.~\ref{eq:ampeqs} is similar to the phase evolution of a single wave in the presence of drag (Eq. 9 of \cite{LestzDuartePoP2021}). Consequently, if there were regimes where a quasi-steady solution was found for both $A_{1}$ and $A_{2}$, then the long time behavior of each would be to have $A_j(t) \rightarrow A_{j,sat} \exp[i t \mathrm{Im}(b_j)/\mathrm{Re}(b_j) ]$, where the exponential term represents a  frequency shift induced by the second mode.

\subsection{Further reduction of amplitude evolution to a single Liénard equation}

The two first-order differential equations for $|\hat{A_j}|$ (Eq.~\ref{eq:ampeqs}) can be combined into a single second-order differential equation where only one of the amplitudes appears. To simplify the resulting expression, we redefine the function to be solved as $\Psi \equiv |\hat{A_2}|^{-2b_0/\mathrm{Re}(b_2)}$ in order to linearize the highest-order derivative. The resulting decoupled equation \footnote{A similar procedure can be done to obtain a second order differential equation in terms of the first wave amplitude. The indices for the second wave should then be switched with the first wave to arrive at this new equation.} is

\begin{widetext}
\begin{equation} \label{eq:secondorderform}
    \begin{aligned}
        &\Psi'' + \frac{2}{\mathrm{Re}(b_2)} \left[\left(b_0\mathrm{Re}(b_2) + \mathrm{Re}(b_1)\mathrm{Re}(b_2) - 2b_0^2\right)\Psi^{-\frac{\mathrm{Re}(b_2)}{b_0}} + \left(2b_0 - \mathrm{Re}(b_2)\right)\right]\Psi' \\
        & \qquad + \frac{4b_0}{\mathrm{Re}(b_2)^2} \left(1 - b_0\Psi^{-\frac{\mathrm{Re}(b_2)}{b_0}}\right)\left[\left(\mathrm{Re}(b_1)\mathrm{Re}(b_2) - b_0^2\right)\Psi^{-\frac{\mathrm{Re}(b_2)}{b_0}}+ \left(b_0 - \mathrm{Re}(b_2)\right)\right] \Psi = 0.
    \end{aligned}
\end{equation}

\end{widetext}

This form provides for a better comparison with textbook tables of equations that allow for analytic solutions. Incidentally, Eq. \ref{eq:secondorderform} has the form of a Li\'{e}nard equation, which can be analytically solvable for certain combinations of $\mathrm{Re}(b_1)$ and $\mathrm{Re}(b_2)$. Two such cases are presented in Appendix B.
For sufficiently small amplitudes, Eq. \ref{eq:secondorderform} can be approximately reduced to a damped linear oscillator equation for $\Psi$ and $|\hat{A_j}|\sim e^t$ is obtained, as expected from Eq. \ref{eq:ampeqs}.

\subsection{Analytical solution for the case of a dominant mode}

Consider the case in which one wave, for instance wave 2, grows independently of wave 1, but wave 1 is still affected by wave 2. Formally, this corresponds to $\mathrm{Re}(b_2)\ll b_0,\mathrm{Re}(b_1)$. Then, its wave amplitude evolution is described by \cite{duarte2018analytical}

\begin{equation}
    |\hat{A_2}|(t) = \frac{|\hat{A_2}|(0)e^t}{\sqrt{1-b_0|\hat{A_2}|^2(0)\left(1-e^{2t}\right)}}.
    \label{eq:A2indep}
    \end{equation}
    
When inserting Eq. \ref{eq:A2indep} into the amplitude evolution for wave 1 [Eq. \ref{eq:ampeqs}], a Stuart-Landau equation with varying coefficients is obtained. This resulting equation becomes linear when rewritten in terms of $|\hat{A_1}|^{-2}$, 

\begin{equation}
    \frac{\mathrm{d}|\hat{A_1}|^{-2}}{\mathrm{d}t} = -2 |\hat{A_1}|^{-2} \left(1 - \mathrm{Re}(b_1) |\hat{A_2}|^2(t) \right) + 2b_0.
\end{equation}
which has the solution

\begin{equation}
    |\hat{A_1}|^{-2}(t) = |\hat{A_1}|^{-2}(0) e^{-2h(t)} + 2b_0\int^t_0e^{-2\left(h(t) - h(t')\right)}dt'
\end{equation}
where
\begin{equation}
h(t) \equiv t - \frac{\mathrm{Re}(b_1)}{2b_0}\log\left|1 - b_0 |\hat{A_2}|^2(0)\left(1- e^{2t}\right)\right|.
\end{equation}
Upon further integration, the final closed-form solution is found to be


\begin{widetext}

\begin{equation}
    |\hat{A_1}|(t) = 
    \begin{cases}
    |\hat{A_1}|(0)e^{h(t)}\left(1 + \frac{|\hat{A_1}|^{2}(0)}{|\hat{A_2}|^{2}(0)} \left(\frac{b_0}{b_0 -\mathrm{Re}(b_1)}\right)\left\{\left[1 -b_0|\hat{A_2}|^2(0)\left(1- e^{2t}\right)\right]^{1-\frac{\mathrm{Re}(b_1)}{b_0}}-1\right\}\right)^{-1/2} & \text{if } \mathrm{Re}(b_1) \neq b_0 \\
    |\hat{A_1}|(0)e^{h(t)}\left\{1 + \frac{|\hat{A_1}|^{2}(0)}{|\hat{A_2}|^{2}(0)} \log\left[1 - b_0 |\hat{A_2}|^2(0)\left(1- e^{2t}\right)\right]\right\}^{-1/2} & \text{if } \mathrm{Re}(b_1) = b_0. \\
    \end{cases}
    \label{eq:analyticssol}
\end{equation}
\end{widetext}

\begin{figure}[th!]
    \centering
    \includegraphics[width = \linewidth]{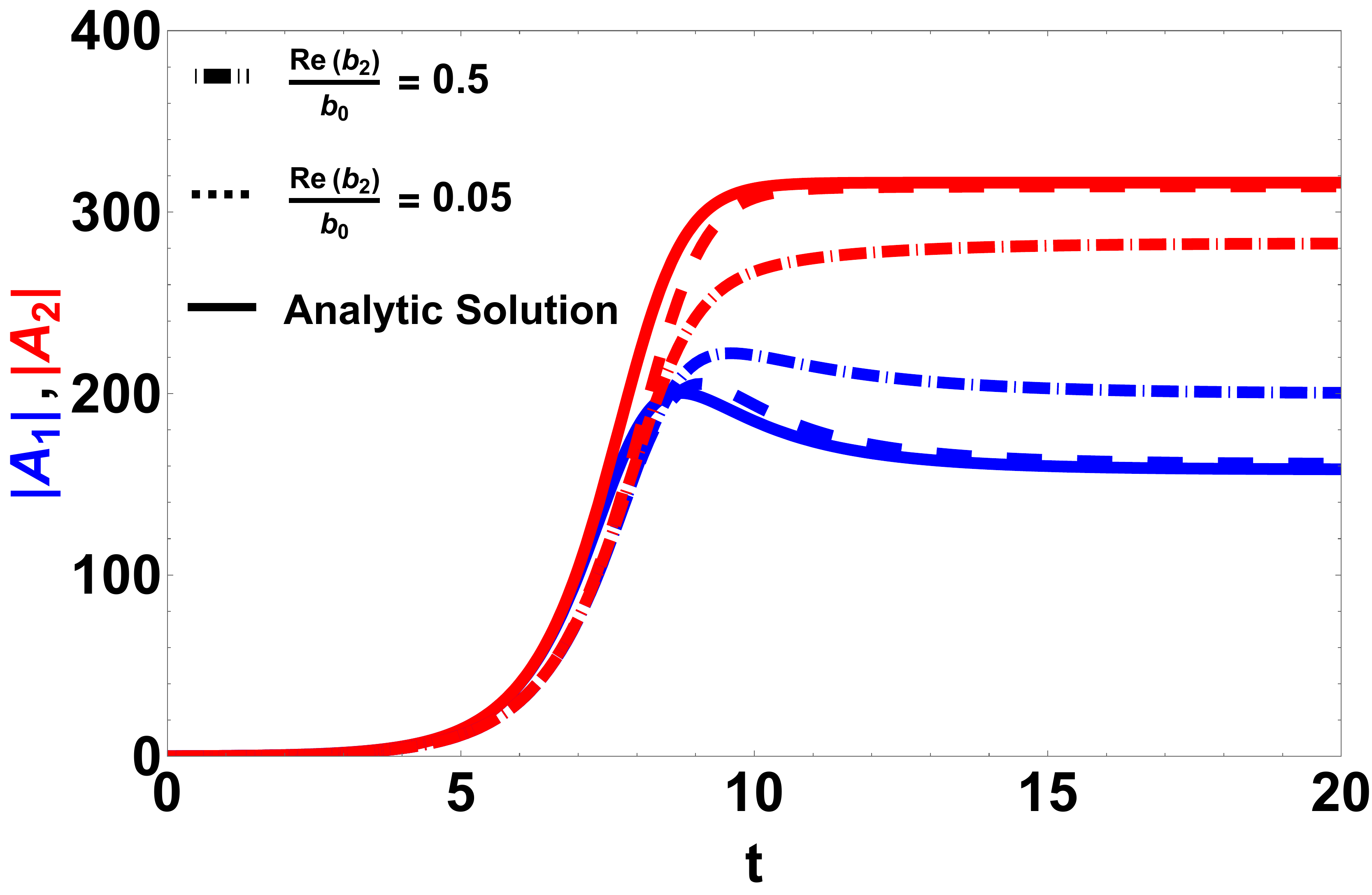}
    \caption{Comparison between the analytic solution (Eq. \ref{eq:analyticssol}) and simulations of Eq. \ref{eq:ampeqs} for the case in which mode 2 regulates the evolution of the mode 1 but mode 1 does not reciprocate back. For this comparison, $\mathrm{Re}(b_1)/b_0 = 0.75$ and $b_0 = 10^{-5}$ are used. The two dashed curves represent simulation data for decreasing values of $\mathrm{Re}(b_2)/b_0$, explicitly $\mathrm{Re}(b_2)/b_0 = 0.5 \ \text{and} \ 0.05$. As the value of $\mathrm{Re}(b_2)/b_0$ decreases to zero, the solution limit for the wave amplitude evolution tends to match the analytic expression in Eq. \ref{eq:analyticssol}. }
    \label{fig:analyticsol}
\end{figure}

For small $t$, $h(t) \sim t $ and Eq. \ref{eq:analyticssol} leads to $|\hat{A_1}|(t) =  |\hat{A_1}|(0)e^{t}$. Fig. \ref{fig:analyticsol} shows that as the interaction term approaches zero, the resulting wave evolution equation approaches the form described in Eq. \ref{eq:analyticssol}. Additionally, when neither wave interacts with the other (i.e., by demanding the additional condition $\mathrm{Re}(b_1)/b_0 \rightarrow 0$), Eq. \ref{eq:analyticssol} recovers Eq. \ref{eq:A2indep}, a case in which both waves saturate independently from each other.

In this same regime, in which two waves saturate independently, the phase evolution is solved by plugging in Eq. \ref{eq:A2indep} to the phase evolution in Eq. \ref{eq:ampeqs}. This results in a closed form for the phase evolution of two independently saturating waves, which is given by 
\begin{equation}
    \begin{aligned}
        \phi_1(t) &= \phi_1(0) - \frac{\mathrm{Im}(b_1)}{2b_0}\log\left[1-b_0|\hat{A_2}|^2(0)\left(1-e^{2t}\right)\right] \\
        \phi_2(t) &= \phi_2(0) - \frac{\mathrm{Im}(b_2)}{2b_0}\log\left[1-b_0|\hat{A_1}|^2(0)\left(1-e^{2t}\right)\right]. \\
    \end{aligned}
    \label{eq:phaseindependent}
\end{equation}
 In the long term limit, the phase velocity for the two waves approach $\phi'_1\rightarrow -\mathrm{Im}(b_1)/b_0$ and $\phi'_2\rightarrow -\mathrm{Im}(b_2)/b_0$, which suggests that, in this limit, the phase growth is linear.

Incidentally, determining the separation of resonances in phase space from $b_1$ and $b_2$ alone is impossible. For given values of $b_1$ and $b_2$, one can uniquely determine $u_1$, $u_2$, $p_1$ and $p_2$. From these,  one can find $k_1$ and $k_2$, as well as $\omega_{1}/\gamma$ and $\omega_{2}/\gamma$. This implies that only for a given $\gamma$, the exact separation between the two resonances can then be determined.

\section{Saturation Levels and Stability Analysis}

Let us analyze properties of the autonomous first-order coupled differential equations (Eq.~\ref{eq:ampeqs}) by defining the functions $f$ and $g$ such that

\begin{equation}
    \begin{aligned}
        & f(|\hat{A_1}|,|\hat{A_2}|) \equiv\frac{\mathrm{d}|\hat{A_1}|}{\mathrm{d}t} = |\hat{A_1}|(1 - b_0|\hat{A_1}|^2 - \mathrm{Re}(b_1)|\hat{A_2}|^2), \\
        & g(|\hat{A_1}|,|\hat{A_2}|) \equiv\frac{\mathrm{d}|\hat{A_2|}}{\mathrm{d}t} = |\hat{A_2}|(1 - b_0|\hat{A_2}|^2 - \mathrm{Re}(b_2)|\hat{A_1}|^2). \\
    \end{aligned}
    \label{eq:diffeq}
\end{equation}

Since the amplitudes are restricted to be positive, there exists up to four fixed points in the system: $(|\hat{A}_1|^F,|\hat{A}_2|^F) = \{ (0,0), (b_0^{-1/2}, 0), (0, b_0^{-1/2}), (|\hat{A_1}|_{sat},|\hat{A_2}|_{sat})\}$ where 
\begin{equation}
\begin{aligned}
|\hat{A_1}|_{sat} =& \sqrt{\frac{b_0-\mathrm{Re}(b_1)}{b_0^2 -\mathrm{Re}(b_1)\mathrm{Re}(b_2)}},
\\
|\hat{A_2}|_{sat} =& \sqrt{\frac{b_0-\mathrm{Re}(b_2)}{b_0^2 -\mathrm{Re}(b_1)\mathrm{Re}(b_2)}}.
\end{aligned}
 \label{eq:satlevels}
\end{equation}

The fourth fixed point is not guaranteed to exist as it can be imaginary. When the argument of the square root is negative for either the first or the second wave, the system then carries only three fixed points.

\begin{table*}[th!]
\renewcommand{\arraystretch}{1.6}
\caption{ Stability of Fixed Points \label{table:results1}}
\begin{ruledtabular}
\begin{tabular}{|c||ccc|}
$(|\hat{A_1}|^F,|\hat{A_2}|^F)$ & $\lambda_1$ & $\lambda_2$ & Stability Type \\
\hline \hline
$(0,0)$ & 1 & 1 & Unstable \\
$(b_0^{-1/2}, 0)$ & $1-\frac{\mathrm{Re}(b_2)}{b_0}$ & -2 & Stable, Saddle Point \\
$(0, b_0^{-1/2})$ & $1-\frac{\mathrm{Re}(b_1)}{b_0}$ & -2 & Stable, Saddle Point \\
$(|\hat{A_1}|_{sat},|\hat{A_2}|_{sat})$ & $-2\frac{(b_0 - \mathrm{Re}(b_1))(b_0 - \mathrm{Re}(b_2)) }{b_0^2-\mathrm{Re}(b_1)\mathrm{Re}(b_2)}$ & -2 & Stable, Saddle Point \\
\end{tabular}
\end{ruledtabular}
\end{table*}

\begin{table*}[th!]
\centering
\renewcommand{\arraystretch}{1.6}
\caption{Region Behavior Classification \label{table:results2}}
\begin{ruledtabular}
\begin{tabular}{|c||ccc|}
Region & Regime & Stable & Saddle Points \\
\hline \hline
A & $\mathrm{Re}(b_1)> b_0$, $\mathrm{Re}(b_2)> b_0$  & $(b_0^{-1/2}, 0)$, $(0,b_0^{-1/2})$ & $(|\hat{A_1}|_{sat},|\hat{A_2}|_{sat})$ \\
B & $\mathrm{Re}(b_1)< b_0$, $\mathrm{Re}(b_2)< b_0$, $\mathrm{Re}(b_1)\mathrm{Re}(b_2)< b_0^2$  & $(|\hat{A_1}|_{sat},|\hat{A_2}|_{sat})$ & $(b_0^{-1/2}, 0)$, $(0,b_0^{-1/2})$ \\
C & $\mathrm{Re}(b_1) > b_0$, $\mathrm{Re}(b_2) < b_0$ & $(0,b_0^{-1/2})$ & $(b_0^{-1/2}, 0)$ \\
D & $\mathrm{Re}(b_1) < b_0$, $\mathrm{Re}(b_2) > b_0$ & $(b_0^{-1/2},0)$ & $(0,b_0^{-1/2})$ \\
E & $\mathrm{Re}(b_1)<0$, $\mathrm{Re}(b_2)<0$, $\mathrm{Re}(b_1)\mathrm{Re}(b_2) > b_0^2$ & - & $(0,b_0^{-1/2})$,  $(b_0^{-1/2}, 0)$ \\
\end{tabular}
\end{ruledtabular}
\end{table*}

By definition, at the fixed points, the amplitudes do not change and $f=g=0$. To analyze the dynamical behavior around them, the time derivative of the wave amplitudes can be approximated using a linear expansion. Therefore, perturbations around fixed points can be treated as though they are centered around the origin instead. This implies that $f(|\hat{A}_1|^F + \delta |\hat{A}_1|, |\hat{A}_2|^F + \delta |\hat{A}_2|) = \frac{\mathrm{d}}{\mathrm{d}t}(\delta|{\hat{A}_1}|)$ and $g(|\hat{A}_1|^F + \delta |\hat{A}_1|, |\hat{A}_2|^F + \delta |\hat{A}_2|) = \frac{\mathrm{d}}{\mathrm{d}t}(\delta|\hat{A_2}|)$. As a result, the linear expansion becomes 
\begin{equation}
    \begin{aligned}
        \frac{\mathrm{d}}{\mathrm{d}t}(\delta|\hat{A_1}|) &\approx \left.\frac{\partial f}{\partial |\hat{A_1}|}\right|_F \delta |\hat{A_1}| + \left.\frac{\partial f}{\partial |\hat{A_2}|}\right|_F \delta |\hat{A_2}|,\\
        \frac{\mathrm{d}}{\mathrm{d}t}(\delta|\hat{A_2}|) &\approx \left.\frac{\partial g}{\partial |\hat{A_1}|}\right|_F \delta |\hat{A_1}| + \left.\frac{\partial g}{\partial |\hat{A_2}|}\right|_F \delta |\hat{A_2}|.\\
    \end{aligned}
    \label{eq:linearization}
\end{equation}

The linearization can also be written in matrix form as $\frac{\mathrm{d}\mathbf{X}}{\mathrm{d}t}= \mathcal{A}\mathbf{X}$ with $\mathbf{X} = \begin{bmatrix} 
    \delta|\hat{A_1}| \\ \delta|\hat{A_2}|
    \end{bmatrix}$ where $\mathcal{A}$ is the Jacobian of the dynamical system defined by
\begin{widetext}
\begin{equation} \label{eq:Jacobian}
    \begin{aligned}
    \mathcal{A} =&
    \begin{bmatrix}
        \left.\frac{\partial f}{\partial |\hat{A_1}|}\right|_F & \left.\frac{\partial f}{\partial |\hat{A_2}|}\right|_F \\
        \left.\frac{\partial g}{\partial |\hat{A_1}|}\right|_F & \left.\frac{\partial g}{\partial |\hat{A_2}|}\right|_F \\
    \end{bmatrix} 
    =
    \begin{pmatrix}
        1 -3b_0(|\hat{A_1}|^F)^2 -\mathrm{Re}(b_1)(|\hat{A_2}|^F)^2 & -2\mathrm{Re}(b_1)|\hat{A_1}|^F|\hat{A_2}|^F\\
        -2\mathrm{Re}(b_2)|\hat{A_1}|^F|\hat{A_2}|^F &  1 -3b_0(|\hat{A_2}|^F)^2 -\mathrm{Re}(b_2)(|\hat{A_1}|^F)^2 \\
    \end{pmatrix}. \\
    \end{aligned}
\end{equation}
\end{widetext}


The general solution for this form is $\mathbf{X} = C_1 \mathbf{v}_1e^{\lambda_1t} + C_2\mathbf{v}_2e^{\lambda_2t}$, where $\mathbf{v}_j$ and $\lambda_j$ are the eigenvectors and eigenvalues of matrix $\mathcal{A}$ respectively. This is an appropriate solution for when the eigenvalues of the linearization are not repeated. A different general form is required otherwise. 

Per the Linearization Theorem, the local topological behavior of the dynamical system around its hyperbolic fixed points is preserved under linearization \cite{wiggins2003introduction}. A fixed point is classified as hyperbolic if the Jacobian of the dynamical system evaluated at the fixed point has no eigenvalues with $\mathrm{Re}(\lambda_j) = 0$.

\begin{figure*}[pt!] 
  \centering
  \includegraphics[width = \linewidth]{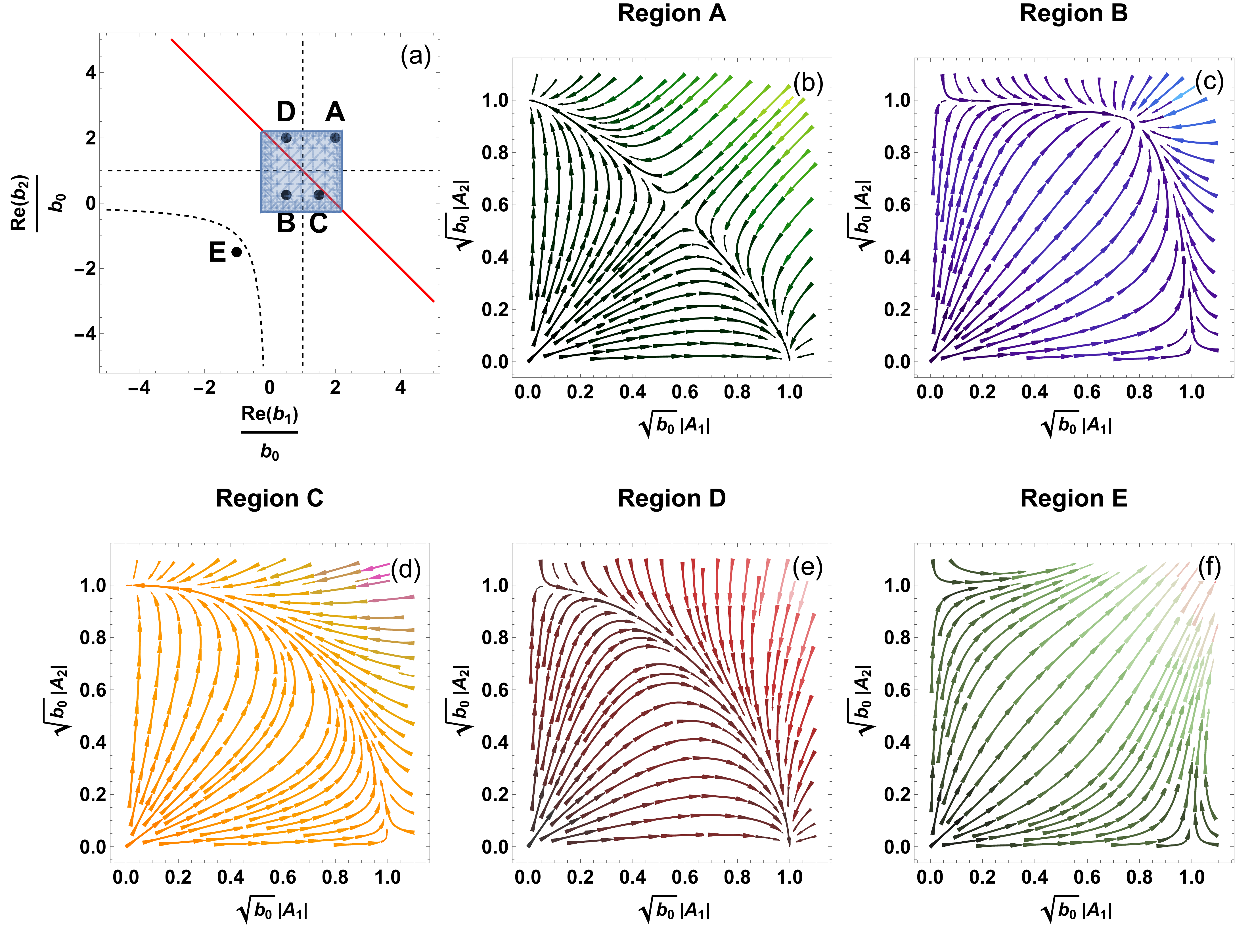}
  \caption{(a) Three bifurcation curves (black dashed lines) divide the parameter space into five distinct regions. The blue filled square represents the acceptable range of the ratios of constants under the assumption that $|u_j|\ll 1$. Specifically, in this graph, $|u_j| < 0.1$ which produces a parameter range of $-0.27b_0 < \mathrm{Re}(b_j) < 2.2b_0$. This blue box spans four regions, namely (b) Region A, (c) Region B, (d) Region C, and (e) Region D. This suggests that there exists four possible wave behaviors that emerge from the reduced differential equations. The solid red curve represents the analytical solution line where constants match the equation of Li\'{e}nard's original form in Appendix B. The line intersects the box of acceptable constants. Sample stream plots are derived for each of the five bifurcation zones using parameter values described by the markers in Fig \ref{fig:stream}(a). From initial conditions, the wave amplitudes evolve parametrically in time following a streamline governed by the vector field set by Eq. \ref{eq:ampeqs}(a,c). For all stream plots, there exists at least three fixed points: the origin which is unstable, and two fixed points that lie on the $|\hat{A_1}|,|\hat{A_2}|$ axes. In Regions A and B, there exists an additional saturation fixed point, which contains nonzero components for $|\hat{A_1}|$ and $|\hat{A_2}|$. In Region A, the saturation fixed point is characterized as a saddle point. Alternately, in Region B, the saturation fixed point is classified as a stable point which allows both modes to saturate. All regions except (f), in Region E, are physical wave behaviors.}
  \label{fig:stream}
\end{figure*}

The signs of the two eigenvalues, determined by the relative magnitude of the constants, determine the classification of the fixed point. Table~\ref{table:results1} provides an overview of the calculated eigenvalues for each fixed point. The behavior of the system is governed by the eigenvalues of the linearized map around the fixed points. Besides the origin, which is a source, all other fixed points have one eigenvalue of value $\lambda_2=-2$, reminiscent of the single mode evolution eigenvalue reported in Ref. \cite{duarte2018analytical}. This requires that these fixed points be either stable or saddle points. Additionally, there exists no oscillatory solutions because there are no eigenvalues that are complex.

Bifurcation analysis dictates that shifting the parameters $\mathrm{Re}(b_1)/b_0$ and $\mathrm{Re}(b_2)/b_0$ causes a change in the signs of the eigenvalues of the linearized system. This variation in parameter space in turn causes a behavioral change of the fixed points of the system. The bifurcation map is divided by three curves and separated into five distinct regions (Fig.~\ref{fig:stream}(a)). The bifurcation lines correspond to $\mathrm{Re}(b_1) = b_0$, $\mathrm{Re}(b_2) = b_0$, and $\mathrm{Re}(b_1)\mathrm{Re}(b_2) = b_0^2$. Table~\ref{table:results2} fully describes the topological behavior for all of the fixed points for each region in Fig.~\ref{fig:stream}(a). For all regions, the origin is always an unstable point. When the magnitudes of the two waves are very small, the amplitude evolution equation (Eq.~\ref{eq:ampeqs}(a,c)) closely resembles an exponential growth equation for each of the wave amplitudes. 

Eqs.~\ref{amp1} and \ref{amp2} are constrained by $|u_j|\ll1$. The complete range for the parameters, $\mathrm{Re}(b_1)/b_0$ and $\mathrm{Re}(b_2)/b_0$, is bounded by the case when $|u_j| = 1$. The ranges of the ratio of the constants can be evaluated from Fig.~\ref{fig:rangeplot} and placed directly onto the bifurcation map. The shaded blue boxed region in Figure~\ref{fig:stream}(a) encompasses the entire range of parameters that are allowable under the constraint $|u_j|\ll1$. Specifically, we see that when the constraint is tightened to $|u_j| < 0.1$, the range of constants, $\mathrm{Re}(b_j)$, extends from $-0.27b_0$ to $2.2b_0$, thereby spanning over four distinct bifurcations regions, A-D. This suggests that aside from behaviors on the bifurcation lines there exists four kinds of dynamics that can occur. Region E does not lie within the acceptable parameter bounds. Figs.~\ref{fig:stream}(b)-(f) describe example stream plots from each of studied regions. For the stream plot in Region E, both waves are observed to grow infinitely, which is physically unachievable and a consequence of the linearization applied in Eq.~\ref{eq:linearization}. Furthermore, combinations of constants that allow for analytical solutions of Li\'{e}nard's equation (see Appendix B) are displayed by the red line in Fig.~\ref{fig:stream}(a). 

In Region A, the two waves initially approach the double saturation saddle point but eventually converge to one of two sinks on $|\hat{A_1}|,|\hat{A_2}|$ space. Depending on the initial conditions one wave dominates over the other. Mode dominance is entirely dependent on the relative magnitude of the interaction terms and the resulting stream plot structure. Unlike other regions, Region A is the only region in which initial conditions affect the long term behavior of the system. In Region B, the two waves will approach the double saturation stable point. Alternately, the fixed points that correlate to one wave engulfing the other wave have changed their behavior to saddle points. In Regions C and D, the double saturation fixed point disappears, and it is seen that one wave dominates over the other regardless of the initial conditions posed. Wave 2 dominates in Region C while wave 1 dominates in Region D. As previously stated, Region E is not within valid parameter bounds, but is shown in this paper for completeness. There also exists other specific cases where the choice of parameters produces points that lie exactly on a bifurcation curve. These cases cannot be ordinarily analyzed through the Linearization Theorem as these fixed points are typically not hyperbolic \cite{wiggins2003introduction}.

\begin{figure}[hb!]
    \centering
    \hspace{-0.9cm}
    \includegraphics[width = 1\linewidth]{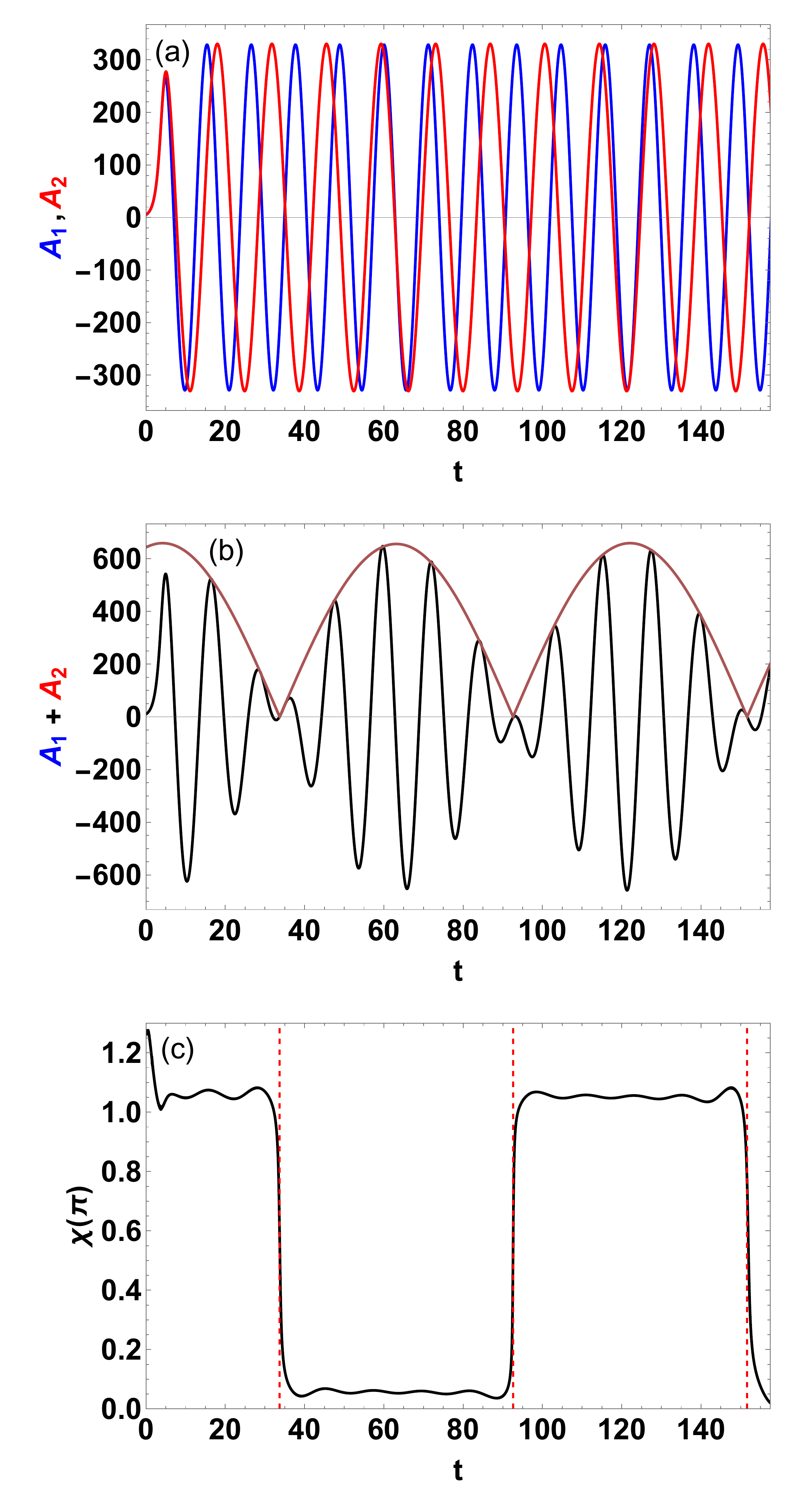}
    \caption{(a) Two waves saturate to amplitudes near each other and are allowed to oscillate by Eqs. \ref{eq:ampeqs}. (b) The amplitudes are combined and beating is observed. A solid brown curve represents the wave envelope $G$. (c) shows the combined phase $\chi$ over a period of time. The red dashed lines represent where phase jumps should occur, according to Eq. \ref{eq:PhaseJumpMasterEq}. In order to calculate $\chi$, a constant value of $\theta$ was subtracted from the argument of the cosine of the combined signal. This explains the initial offset in part (c), before saturation was achieved at around $t=9$ (see Fig. \ref{fig:doublesat}(b)). Apart from this initial phase drift, all other jumps occur by a factor of nearly $\pi$. }
    \label{fig:phasejump}
\end{figure}

\section{Phase jumps and amplitude beatings}

The system of coupled evolution equations treated in this paper (Eqs. \ref{eq:ampeqs}) are found to lead to beating and phase jumps whenever the two amplitudes saturate at similar levels. Analytically, this phenomenon can be understood as follows. Consider the interference between two signals of similar amplitudes $F$ and $(1+\epsilon)F$ (with $\epsilon\ll1$) with distinct phases,

\begin{equation}
F\left[\left(1+\epsilon\right)\cos\left(\theta+\delta\theta\right)+\cos\left(\theta-\delta\theta\right)\right]=G\cos\left(\theta+\chi\right)\label{eq:PhaseJumpMasterEq}
\end{equation}

Since Region B is the only zone that supports double wave saturation, beating and phase jumps are exclusive to this region. The combined signal can be written using the new amplitude $G$ and the new phase $\chi$, as shown in Eq. \ref{eq:PhaseJumpMasterEq}, where 

$$
\chi=\arctan\left[\frac{\epsilon\tan\delta\theta}{2+\epsilon}\right];\quad G=F\sqrt{\epsilon^{2}+4\left(\cos^{2}\delta\theta+\epsilon\cos\delta\theta\right)}.
$$
The changes in $\chi$ around $\delta\theta=\left(n+1/2\right)\pi$ (with $n\in\mathbb{Z}$) occur in jumps of $\pi-\epsilon$.

At saturation (with amplitude levels given by Eq. \ref{eq:satlevels}), the signal of the two combined oscillations given by Eqs. \ref{eq:ampeqs} is 
$$
|\hat{A_{1}}|_{sat}\cos\left[\text{Im}\left(b_{1}\right)|\hat{A_{2}}|^{2}_{sat}t\right]+|\hat{A_{2}}|_{sat}\cos\left[\text{Im}\left(b_{2}\right)|\hat{A_{1}}|^{2}_{sat}t\right],
$$ which can be cast in the form of Eq. \ref{eq:PhaseJumpMasterEq} by defining
$$
\theta\equiv\frac{\text{Im}\left(b_{1}\right)|\hat{A_{2}}|^{2}_{sat}t+\text{Im}\left(b_{2}\right)|\hat{A_{1}}|^{2}_{sat}t}{2}
$$ and
$$
\delta\theta\equiv\frac{\text{Im}\left(b_{1}\right)|\hat{A_{2}}|^{2}_{sat}t-\text{Im}\left(b_{2}\right)|\hat{A_{1}}|^{2}_{sat}t}{2}
$$
provided that $||\hat{A_{1}}|_{sat}-|\hat{A_{2}}|_{sat}|\ll|\hat{A_{1}}|_{sat},|\hat{A_{2}}|_{sat}$.



The phase jump phenomenon is illustrated in Fig. \ref{fig:phasejump} by solving Eq. \ref{eq:ampeqs} for a case in which their amplitudes saturate at comparable levels. When the signals are combined, the resulting waves exhibit beating and periodic amplitude crashing. The combined signal plot shows phase jumps of magnitude of approximately $\pi$ that occur when the wave envelope amplitude $G$ approaches zero. Fig. \ref{fig:phasejump} shows that if there are two waves, bursting can happen in this system even when $\nu/\gamma \gg 1$ -- a regime in which only quasi-steady behavior exists for a single wave. As the underlying perturbed distribution is expected to depend on the combined action of the two amplitudes, this beating can lead to intermittent, bursty transport at the beat frequency, which has implications for peak heat load increase.

Phase jumps are common in the dynamics of coupled oscillators \cite{LeeKwakLim1998}. Phase jumps of Alfvén eigenmodes have been observed in fusion experiments and interpreted as a signature of chaotic behavior \cite{HeeterPRL2000}.  More recently, Bierwage \cite{BierwageNF2017,bierwage2021effect} performed detailed studies on the effect of phase jumps and amplitude beating in connection with frequency chirping. In our system, such chirping does not occur since the phase space is stochasticized around the resonances due to high $\hat{\nu}$ and therefore it does not support propagating coherent structures. The beating and phase jumps are expected to have no significant effect on the present dynamics, since phase space structures are already scrambled by collisions. Hence, the beating and the associated phase jumps are assumed to be just a passive byproduct with no active role in the dynamics in the present case.

The Kuramoto order parameter \cite{kuramoto1975self,acebron2005kuramoto}, which describes the synchronization for many coupled oscillators, can be applied to our binary oscillating system similar to Fig.~\ref{fig:phasejump}(a). The Kuramoto order parameter for two oscillators takes the form 

\begin{equation}
    Z = r(t) e^{i\psi(t)} = \frac{1}{2}\left(e^{i\phi_1(t)} + e^{i\phi_2(t)}\right)
    \label{eq:Kuramoto}
\end{equation}
where $r(t) = \cos\left(\delta\theta(t)\right)$ is the phase coherence and $\psi(t) = \theta(t)$ is the average phase.

It is also possible to analyze synchronicity conditions for two waves. They are considered synchronous if $|\phi_1(t) - \phi_2(t)| < C$, where $C$ is a constant. This requires that at the long term limit the difference of the time derivatives of the amplitude phase must tend to zero. If the difference is not exactly zero, then the phase of the two amplitudes will drift apart at long periods of time. The synchronicity condition only applies to waves in Region B and leads to

\begin{equation} \label{eq:synchro}
    \frac{\mathrm{Im}(b_1)/b_0}{\mathrm{Im}(b_2)/b_0} = \frac{1 - \mathrm{Re}(b_1)/b_0}{1 -\mathrm{Re}(b_2)/b_0} = \frac{|\hat{A_1}|_{sat}^ 2}{|\hat{A_2}|_{sat}^2}.
\end{equation}
In the other regions, the synchronicity condition fails to be achieved. 

\section{Comparison between the reduced time-local and the time-delayed systems }

For situations in which steady saturation exists, the reduced model described by Eq. \ref{eq:ampeqs} is is found to be in fair agreement with the more complex system in which time-delays are retained (Eqs. \ref{amp1} and \ref{amp2}). The numerical methodology for solving the latter is based on Ref. \cite{HeeterThesis} and is described in Appendix C.

\begin{figure*}[t]
    \centering
    \includegraphics[width = 0.95\linewidth]{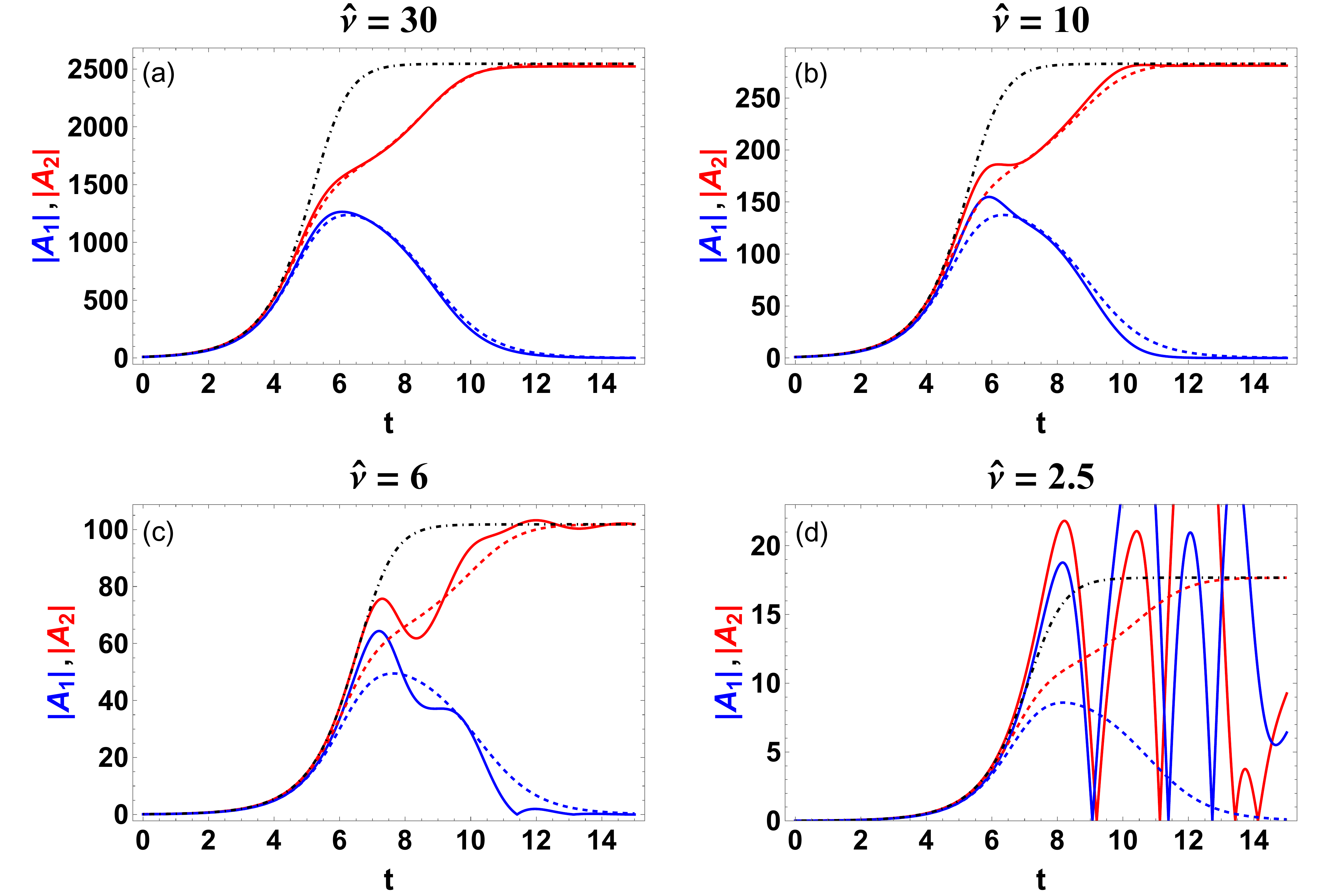}
    \caption{Numerical results for a case in Region A where one wave dominates over the other wave, for (a) $\hat{\nu}=30$, (b) $\hat{\nu}=10$, (c) $\hat{\nu}=6$ and (d) $\hat{\nu}=2.5$. The red and blue curves represent the first and the second wave, respectively. Solid lines represent the numerical solution of the integro-differential Eqs.  \ref{amp1} and \ref{amp2} while dashed lines represent the solution produced by the time-local reduced model of Eq. \ref{eq:ampeqs}. The black dashed line represents a single mode evolution described by Eq. \ref{eq:A2indep}. The presence of mode 1 causes an attenuation in the growth of mode 2. For sufficiently small values of $\hat{\nu}$, the integro-differential equations can lead to unbounded oscillatory growth. Like the double wave saturation scenario, there exists a critical value where the poor memory assumption is not valid. These plots were constructed using two waves with the same wave number and frequency, but have slightly different initial wave amplitudes. For this case specifically, the initially larger wave eventually dominates over the smaller wave for all plots excluding the case where the amplitude blows up. These plots were constructed using $k_1/k_2 = \omega_1/\omega_2 = 1$.}
    \label{fig:singlesat}
\end{figure*}

\begin{figure*}
    \centering
    \includegraphics[width = 0.95\linewidth]{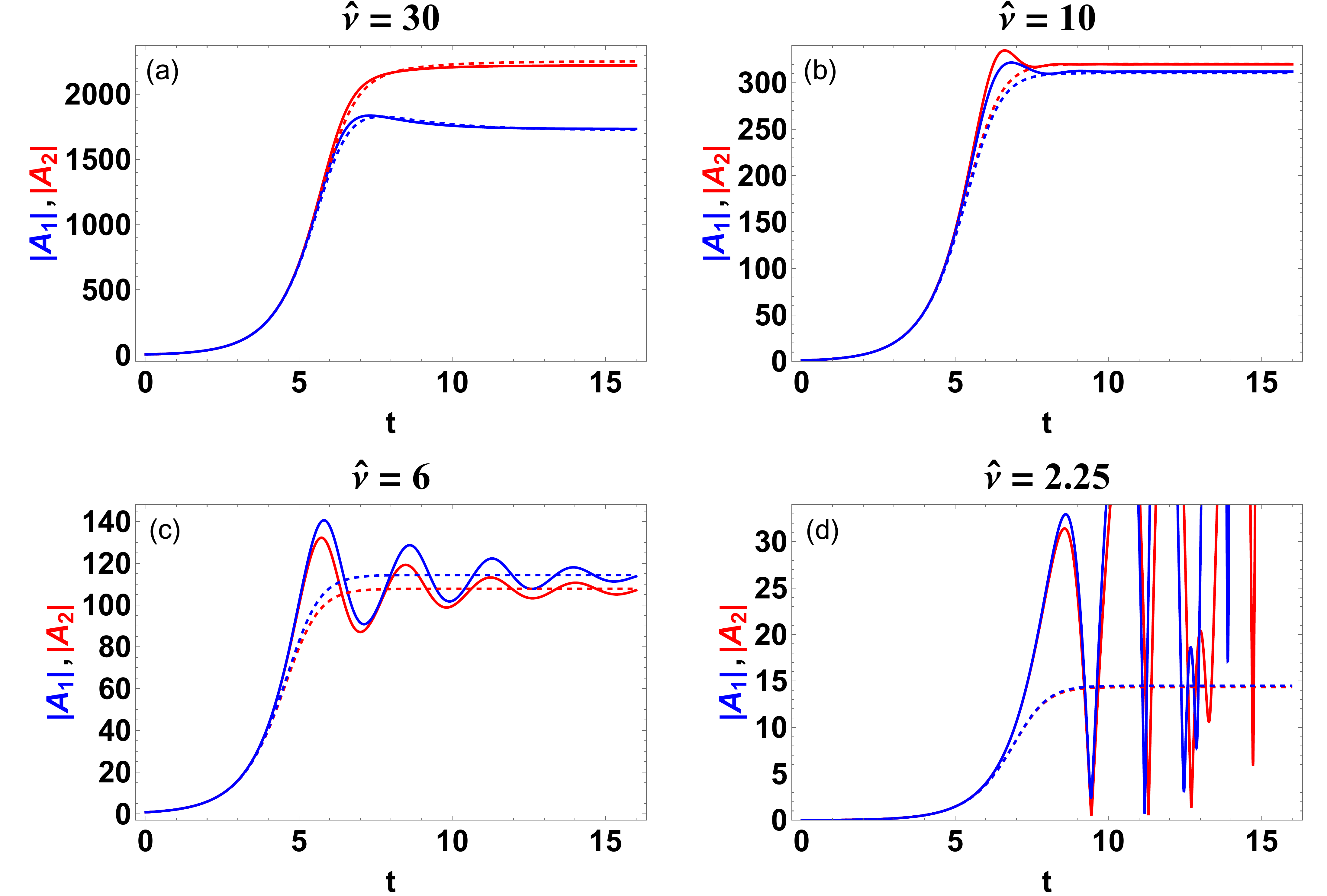}
    \caption{Numerical results with varying $\hat{\nu}$ are presented for a case where both waves saturate to non-zero amplitudes, for (a) $\hat{\nu}=30$, (b) $\hat{\nu}=10$, (c) $\hat{\nu}=6$ and (d) $\hat{\nu}=2.25$. The red and blue solid lines represent the numerical solution of Eq. \ref{amp1} and \ref{amp2}. The red and blue dashed lines represent the reduced model in Eq. \ref{eq:ampeqs} using the poor memory argument for simplification. At large values of $\hat{\nu}$, there is large agreement between the original cubic equation and the reduced analytic model. However, with smaller values of $\hat{\nu}$, oscillatory behavior starts to appears around the saturation level. At even lower values of $\hat{\nu}$, the amplitude blows up. These plots were constructed using $k_1/k_2 = 1.67$ and $\omega_1/\omega_2 = 1.20$. Both waves start at the same amplitude.}
    \label{fig:doublesat}
\end{figure*}

The case in which one wave regulates the growth of the other is shown in Fig. \ref{fig:singlesat} (Region A of Fig. \ref{fig:stream}(a)). The case for double saturation waves is presented in Figure \ref{fig:doublesat} (Region B of Fig. \ref{fig:stream}(a)). In both figures, at large values of $\hat{\nu}$ oscillations are heavily damped and the reduced time-local model agrees well the time-delayed model. At $\hat{\nu}=10$, the time-local system is still able to capture the essence of the dynamics of the time-delayed system, while not being able to capture its fine details. At $\hat{\nu}=6$, the oscillations are more prominent but still dampen to converge to the expected saturated value. Finally, below a critical value of $\hat{\nu}$, the solution of the time-delayed system blows up, which indicates that the perturbative ordering used in its derivation is no longer valid, and the reduced model fails to approximate the wave explosion early trend. For reference, the critical value that separates steady and blow-up solutions for the case of a single wave evolution is $\hat{\nu}=4.38$ \cite{BerkPRL1996}.

The beatings of Fig. \ref{fig:phasejump}(b) and the beatings of Figs. \ref{fig:singlesat}(d) and \ref{fig:doublesat}(d) are very different in nature. In Fig. \ref{fig:phasejump}(b), it beats due to the interference between two waves at similar amplitudes. In Figs. \ref{fig:singlesat}(d) and \ref{fig:doublesat}(d), the ``beating" is actually a strong pulsation or explosion of a single mode amplitude in a strongly driven scenario in which the original ordering $\gamma_L-\gamma_d \ll \gamma_L$ is no longer valid.

Additionally, when compared to the single mode evolution in Eq. \ref{eq:A2indep}, the two-mode interaction exhibits a noticeable delay in saturation speed. At early times before wave saturation in Fig. \ref{fig:singlesat}, the dominant mode adheres to the single mode saturation model, but later deviates due to interaction with the subdominant mode.



Since the original model is calculated with numerical integrals, some error analysis is presented. All numerical simulations are designed to start with small initial amplitudes. However, the rate of change of the amplitudes depends on an integral, so initial conditions will affect the growth of wave trajectories and numerical saturation value.
Another variable that significantly influences the numerical saturation level is the time step value. Generally, increasing the time step size causes a noticeable shift in saturation level during the simulation. To maintain a high degree of accuracy, the time step size should be set as small as allowed computationally. The explicit dependence of the time step size on saturation level is presented in Appendix C. For the simulations in Figs. \ref{fig:singlesat} and \ref{fig:doublesat}, the time step size was set to $\Delta t =0.002$, to achieve appropriate convergence.

\section{Discussion}

The resonant interaction between coupled modes and particles has been studied using collisional kinetic theory. Near marginal stability, the dynamics is governed by coupled nonlinear time-delayed, integro-differential equations \cite{Zalesny}. For experimentally-relevant scenarios in which effective collisions in the resonance layer happen at a timescale shorter than the characteristic wave growth time, those equations have been simplified to time-local, ordinary differential equations with complex coefficients. These reduced equations can be cast into a differential equation of Li\'{e}nard form and be solved analytically for particular cases. Numerical comparisons between the two systems have indicated that the reduced time-local approach is reasonable whenever the modes evolve quasi-steadily to saturation. When the two modes saturate at nearby levels, the total signal exhibits repetitive beatings accompanied by phase jumps of approximately $\pi$ occurs. 

This paper focuses on the application of a Krook-type collision operator. However, without loss of generality, a similar procedure for the reduction of Eqs. \ref{amp1} and \ref{amp2} can be performed for different collision operators, which would generate different coefficients. Future work will apply our procedure to establish a particle transport equation in the presence of closely spaced waves. The present work can be readily applicable to the study of the destabilization of Alfvénic eigenmodes due to fast ions in fusion plasmas. Additionally, the methods developed here can be applied to the self-regulation between modes of distinct nature \cite{LiuPRL2022} and their consequent particle transport.


\section*{Acknowledgements}
The authors thank J. B. Lestz, A. Bierwage, C. Hamilton, I. L. Caldas,  and W. Sengupta for stimulating discussions. This manuscript is based upon work supported by the US Department of Energy, Office of Science, Office of Fusion Energy Sciences, and has been authored by Princeton University under Contract Number DE-AC02-09CH11466 with the US Department of Energy. The publisher, by accepting the article for publication acknowledges, that the United States Government retains a non-exclusive, paid-up, irrevocable, world-wide license to publish or reproduce the published form of this manuscript, or allow others to do so, for United States Government purposes.

\bibliography{bibliography.bib}

\begin{thebibliography}{42}%
\makeatletter
\providecommand \@ifxundefined [1]{%
 \@ifx{#1\undefined}
}%
\providecommand \@ifnum [1]{%
 \ifnum #1\expandafter \@firstoftwo
 \else \expandafter \@secondoftwo
 \fi
}%
\providecommand \@ifx [1]{%
 \ifx #1\expandafter \@firstoftwo
 \else \expandafter \@secondoftwo
 \fi
}%
\providecommand \natexlab [1]{#1}%
\providecommand \enquote  [1]{``#1''}%
\providecommand \bibnamefont  [1]{#1}%
\providecommand \bibfnamefont [1]{#1}%
\providecommand \citenamefont [1]{#1}%
\providecommand \href@noop [0]{\@secondoftwo}%
\providecommand \href [0]{\begingroup \@sanitize@url \@href}%
\providecommand \@href[1]{\@@startlink{#1}\@@href}%
\providecommand \@@href[1]{\endgroup#1\@@endlink}%
\providecommand \@sanitize@url [0]{\catcode `\\12\catcode `\$12\catcode
  `\&12\catcode `\#12\catcode `\^12\catcode `\_12\catcode `\%12\relax}%
\providecommand \@@startlink[1]{}%
\providecommand \@@endlink[0]{}%
\providecommand \url  [0]{\begingroup\@sanitize@url \@url }%
\providecommand \@url [1]{\endgroup\@href {#1}{\urlprefix }}%
\providecommand \urlprefix  [0]{URL }%
\providecommand \Eprint [0]{\href }%
\providecommand \doibase [0]{http://dx.doi.org/}%
\providecommand \selectlanguage [0]{\@gobble}%
\providecommand \bibinfo  [0]{\@secondoftwo}%
\providecommand \bibfield  [0]{\@secondoftwo}%
\providecommand \translation [1]{[#1]}%
\providecommand \BibitemOpen [0]{}%
\providecommand \bibitemStop [0]{}%
\providecommand \bibitemNoStop [0]{.\EOS\space}%
\providecommand \EOS [0]{\spacefactor3000\relax}%
\providecommand \BibitemShut  [1]{\csname bibitem#1\endcsname}%
\let\auto@bib@innerbib\@empty
\bibitem [{\citenamefont {Kadomtsev}(1965)}]{kadomtsev1965plasma}%
  \BibitemOpen
  \bibfield  {author} {\bibinfo {author} {\bibfnamefont {B.~B.}\ \bibnamefont
  {Kadomtsev}},\ }\href@noop {} {\emph {\bibinfo {title} {Plasma Turbulence}}}\
  (\bibinfo  {publisher} {Academic Press},\ \bibinfo {year} {1965})\BibitemShut
  {NoStop}%
\bibitem [{\citenamefont {Diamond}\ \emph {et~al.}(2010)\citenamefont
  {Diamond}, \citenamefont {Itoh},\ and\ \citenamefont
  {Itoh}}]{diamond2010modern}%
  \BibitemOpen
  \bibfield  {author} {\bibinfo {author} {\bibfnamefont {P.~H.}\ \bibnamefont
  {Diamond}}, \bibinfo {author} {\bibfnamefont {S.-I.}\ \bibnamefont {Itoh}}, \
  and\ \bibinfo {author} {\bibfnamefont {K.}~\bibnamefont {Itoh}},\ }\href@noop
  {} {\emph {\bibinfo {title} {Modern Plasma Physics: Volume 1, Physical
  Kinetics of Turbulent Plasmas}}}\ (\bibinfo  {publisher} {Cambridge
  University Press},\ \bibinfo {year} {2010})\BibitemShut {NoStop}%
\bibitem [{\citenamefont {Haas}(2011)}]{haasquantum}%
  \BibitemOpen
  \bibfield  {author} {\bibinfo {author} {\bibfnamefont {F.}~\bibnamefont
  {Haas}},\ }\href@noop {} {\emph {\bibinfo {title} {Quantum Plasmas - An
  Hydrodynamic Approach}}}\ (\bibinfo  {publisher} {Springer},\ \bibinfo {year}
  {2011})\BibitemShut {NoStop}%
\bibitem [{\citenamefont
  {Mikhailovskii}(1998)}]{mikhailovskii2017instabilities}%
  \BibitemOpen
  \bibfield  {author} {\bibinfo {author} {\bibfnamefont {A.~B.}\ \bibnamefont
  {Mikhailovskii}},\ }\href@noop {} {\emph {\bibinfo {title} {Instabilities in
  a Confined Plasma}}}\ (\bibinfo  {publisher} {CRC Press},\ \bibinfo {year}
  {1998})\BibitemShut {NoStop}%
\bibitem [{\citenamefont {O'Neil}(1965)}]{ONiel1965}%
  \BibitemOpen
  \bibfield  {author} {\bibinfo {author} {\bibfnamefont {T.}~\bibnamefont
  {O'Neil}},\ }\href {\doibase http://dx.doi.org/10.1063/1.1761193} {\bibfield
  {journal} {\bibinfo  {journal} {Phys. Fluids}\ }\textbf {\bibinfo {volume}
  {8}},\ \bibinfo {pages} {2255} (\bibinfo {year} {1965})}\BibitemShut
  {NoStop}%
\bibitem [{\citenamefont {Mazitov}(1965)}]{Mazitov1965}%
  \BibitemOpen
  \bibfield  {author} {\bibinfo {author} {\bibfnamefont {R.~K.}\ \bibnamefont
  {Mazitov}},\ }\href {\doibase 10.1007/BF00914365} {\bibfield  {journal}
  {\bibinfo  {journal} {J. Appl. Mech. Tech. Phys.}\ }\textbf {\bibinfo
  {volume} {6}},\ \bibinfo {pages} {22} (\bibinfo {year} {1965})}\BibitemShut
  {NoStop}%
\bibitem [{\citenamefont {Berk}\ and\ \citenamefont
  {Breizman}(1990)}]{BerkBreizman1990a}%
  \BibitemOpen
  \bibfield  {author} {\bibinfo {author} {\bibfnamefont {H.~L.}\ \bibnamefont
  {Berk}}\ and\ \bibinfo {author} {\bibfnamefont {B.~N.}\ \bibnamefont
  {Breizman}},\ }\href {\doibase 10.1063/1.859404} {\bibfield  {journal}
  {\bibinfo  {journal} {Phys. Fluids B}\ }\textbf {\bibinfo {volume} {2}},\
  \bibinfo {pages} {2226} (\bibinfo {year} {1990})}\BibitemShut {NoStop}%
\bibitem [{\citenamefont {Sagdeev}\ and\ \citenamefont
  {Galeev}(1969)}]{sagdeev1969nonlinear}%
  \BibitemOpen
  \bibfield  {author} {\bibinfo {author} {\bibfnamefont {R.~Z.}\ \bibnamefont
  {Sagdeev}}\ and\ \bibinfo {author} {\bibfnamefont {A.~A.}\ \bibnamefont
  {Galeev}},\ }\href@noop {} {\emph {\bibinfo {title} {\textit{Nonlinear Plasma
  Theory}}}}\ (\bibinfo  {publisher} {W. A. Benjamin, Inc},\ \bibinfo {year}
  {1969})\BibitemShut {NoStop}%
\bibitem [{\citenamefont {Stix}(1992)}]{stix1992waves}%
  \BibitemOpen
  \bibfield  {author} {\bibinfo {author} {\bibfnamefont {T.~H.}\ \bibnamefont
  {Stix}},\ }\href@noop {} {\emph {\bibinfo {title} {Waves in plasmas}}}\
  (\bibinfo  {publisher} {Springer Science \& Business Media},\ \bibinfo {year}
  {1992})\BibitemShut {NoStop}%
\bibitem [{\citenamefont {Sagdeev}\ \emph {et~al.}(1988)\citenamefont
  {Sagdeev}, \citenamefont {Usikov},\ and\ \citenamefont
  {Zaslavsky}}]{sagdeev1988nonlinear}%
  \BibitemOpen
  \bibfield  {author} {\bibinfo {author} {\bibfnamefont {R.~Z.}\ \bibnamefont
  {Sagdeev}}, \bibinfo {author} {\bibfnamefont {D.~A.}\ \bibnamefont {Usikov}},
  \ and\ \bibinfo {author} {\bibfnamefont {G.~M.}\ \bibnamefont {Zaslavsky}},\
  }\href@noop {} {\emph {\bibinfo {title} {Nonlinear physics - From the
  pendulum to turbulence and chaos}}}\ (\bibinfo  {publisher} {Harwood Academic
  Publishers},\ \bibinfo {year} {1988})\BibitemShut {NoStop}%
\bibitem [{\citenamefont {Zaleśny}\ \emph {et~al.}(2011)\citenamefont
  {Zaleśny}, \citenamefont {Galant}, \citenamefont {Lisak}, \citenamefont
  {Marczyński}, \citenamefont {Berczyński}, \citenamefont {Gałkowski},\ and\
  \citenamefont {Berczyński}}]{Zalesny}%
  \BibitemOpen
  \bibfield  {author} {\bibinfo {author} {\bibfnamefont {J.}~\bibnamefont
  {Zaleśny}}, \bibinfo {author} {\bibfnamefont {G.}~\bibnamefont {Galant}},
  \bibinfo {author} {\bibfnamefont {M.}~\bibnamefont {Lisak}}, \bibinfo
  {author} {\bibfnamefont {S.}~\bibnamefont {Marczyński}}, \bibinfo {author}
  {\bibfnamefont {P.}~\bibnamefont {Berczyński}}, \bibinfo {author}
  {\bibfnamefont {A.}~\bibnamefont {Gałkowski}}, \ and\ \bibinfo {author}
  {\bibfnamefont {S.}~\bibnamefont {Berczyński}},\ }\href {\doibase
  10.1063/1.3601136} {\bibfield  {journal} {\bibinfo  {journal} {Phys.
  Plasmas}\ }\textbf {\bibinfo {volume} {18}},\ \bibinfo {pages} {062109}
  (\bibinfo {year} {2011})}\BibitemShut {NoStop}%
\bibitem [{\citenamefont {Berk}\ \emph {et~al.}(1996)\citenamefont {Berk},
  \citenamefont {Breizman},\ and\ \citenamefont {Pekker}}]{BerkPRL1996}%
  \BibitemOpen
  \bibfield  {author} {\bibinfo {author} {\bibfnamefont {H.~L.}\ \bibnamefont
  {Berk}}, \bibinfo {author} {\bibfnamefont {B.~N.}\ \bibnamefont {Breizman}},
  \ and\ \bibinfo {author} {\bibfnamefont {M.}~\bibnamefont {Pekker}},\ }\href
  {\doibase 10.1103/PhysRevLett.76.1256} {\bibfield  {journal} {\bibinfo
  {journal} {Phys. Rev. Lett.}\ }\textbf {\bibinfo {volume} {76}},\ \bibinfo
  {pages} {1256} (\bibinfo {year} {1996})}\BibitemShut {NoStop}%
\bibitem [{\citenamefont {Galant}\ \emph {et~al.}(2011)\citenamefont {Galant},
  \citenamefont {Zaleśny}, \citenamefont {Lisak}, \citenamefont
  {Berczyński},\ and\ \citenamefont {Berczyński}}]{Galant_2011}%
  \BibitemOpen
  \bibfield  {author} {\bibinfo {author} {\bibfnamefont {G.}~\bibnamefont
  {Galant}}, \bibinfo {author} {\bibfnamefont {J.}~\bibnamefont {Zaleśny}},
  \bibinfo {author} {\bibfnamefont {M.}~\bibnamefont {Lisak}}, \bibinfo
  {author} {\bibfnamefont {P.}~\bibnamefont {Berczyński}}, \ and\ \bibinfo
  {author} {\bibfnamefont {S.}~\bibnamefont {Berczyński}},\ }\href {\doibase
  10.1088/0031-8949/83/05/055502} {\bibfield  {journal} {\bibinfo  {journal}
  {Phys. Scripta}\ }\textbf {\bibinfo {volume} {83}},\ \bibinfo {pages}
  {055502} (\bibinfo {year} {2011})}\BibitemShut {NoStop}%
\bibitem [{\citenamefont {Galant}(2013)}]{Galant+2013+1598+1604}%
  \BibitemOpen
  \bibfield  {author} {\bibinfo {author} {\bibfnamefont {G.}~\bibnamefont
  {Galant}},\ }\href {\doibase doi:10.2478/s11534-013-0279-0} {\bibfield
  {journal} {\bibinfo  {journal} {Open Phys.}\ }\textbf {\bibinfo {volume}
  {11}},\ \bibinfo {pages} {1598} (\bibinfo {year} {2013})}\BibitemShut
  {NoStop}%
\bibitem [{\citenamefont {Sharma}\ \emph {et~al.}(2012)\citenamefont {Sharma},
  \citenamefont {Dev~Shrimali},\ and\ \citenamefont {Kumar~Dana}}]{Sharma2012}%
  \BibitemOpen
  \bibfield  {author} {\bibinfo {author} {\bibfnamefont {A.}~\bibnamefont
  {Sharma}}, \bibinfo {author} {\bibfnamefont {M.}~\bibnamefont
  {Dev~Shrimali}}, \ and\ \bibinfo {author} {\bibfnamefont {S.}~\bibnamefont
  {Kumar~Dana}},\ }\href {\doibase 10.1063/1.4729459} {\bibfield  {journal}
  {\bibinfo  {journal} {Chaos}\ }\textbf {\bibinfo {volume} {22}},\ \bibinfo
  {pages} {023147} (\bibinfo {year} {2012})}\BibitemShut {NoStop}%
\bibitem [{\citenamefont {Duarte}\ \emph {et~al.}(2019)\citenamefont {Duarte},
  \citenamefont {Gorelenkov}, \citenamefont {White},\ and\ \citenamefont
  {Berk}}]{duarte2019collisional}%
  \BibitemOpen
  \bibfield  {author} {\bibinfo {author} {\bibfnamefont {V.~N.}\ \bibnamefont
  {Duarte}}, \bibinfo {author} {\bibfnamefont {N.~N.}\ \bibnamefont
  {Gorelenkov}}, \bibinfo {author} {\bibfnamefont {R.~B.}\ \bibnamefont
  {White}}, \ and\ \bibinfo {author} {\bibfnamefont {H.~L.}\ \bibnamefont
  {Berk}},\ }\href {\doibase 10.1063/1.5129260} {\bibfield  {journal} {\bibinfo
   {journal} {Phys. Plasmas}\ }\textbf {\bibinfo {volume} {26}},\ \bibinfo
  {pages} {120701} (\bibinfo {year} {2019})}\BibitemShut {NoStop}%
\bibitem [{\citenamefont {Schneller}\ \emph {et~al.}(2015)\citenamefont
  {Schneller}, \citenamefont {Lauber},\ and\ \citenamefont
  {Briguglio}}]{Schneller_2016}%
  \BibitemOpen
  \bibfield  {author} {\bibinfo {author} {\bibfnamefont {M.}~\bibnamefont
  {Schneller}}, \bibinfo {author} {\bibfnamefont {P.}~\bibnamefont {Lauber}}, \
  and\ \bibinfo {author} {\bibfnamefont {S.}~\bibnamefont {Briguglio}},\ }\href
  {\doibase 10.1088/0741-3335/58/1/014019} {\bibfield  {journal} {\bibinfo
  {journal} {Plasma Phys. Control. Fusion}\ }\textbf {\bibinfo {volume} {58}},\
  \bibinfo {pages} {014019} (\bibinfo {year} {2015})}\BibitemShut {NoStop}%
\bibitem [{\citenamefont {Fitzgerald}\ \emph {et~al.}(2016)\citenamefont
  {Fitzgerald}, \citenamefont {Sharapov}, \citenamefont {Rodrigues},\ and\
  \citenamefont {Borba}}]{FitzgeraldITER_NF2016}%
  \BibitemOpen
  \bibfield  {author} {\bibinfo {author} {\bibfnamefont {M.}~\bibnamefont
  {Fitzgerald}}, \bibinfo {author} {\bibfnamefont {S.}~\bibnamefont
  {Sharapov}}, \bibinfo {author} {\bibfnamefont {P.}~\bibnamefont {Rodrigues}},
  \ and\ \bibinfo {author} {\bibfnamefont {D.}~\bibnamefont {Borba}},\ }\href
  {http://stacks.iop.org/0029-5515/56/i=11/a=112010} {\bibfield  {journal}
  {\bibinfo  {journal} {Nuclear Fusion}\ }\textbf {\bibinfo {volume} {56}},\
  \bibinfo {pages} {112010} (\bibinfo {year} {2016})}\BibitemShut {NoStop}%
\bibitem [{\citenamefont {Tagger}\ \emph {et~al.}(1987)\citenamefont {Tagger},
  \citenamefont {Sygnet}, \citenamefont {Athanassoula},\ and\ \citenamefont
  {Pellat}}]{tagger1987nonlinear}%
  \BibitemOpen
  \bibfield  {author} {\bibinfo {author} {\bibfnamefont {M.}~\bibnamefont
  {Tagger}}, \bibinfo {author} {\bibfnamefont {J.}~\bibnamefont {Sygnet}},
  \bibinfo {author} {\bibfnamefont {E.}~\bibnamefont {Athanassoula}}, \ and\
  \bibinfo {author} {\bibfnamefont {R.}~\bibnamefont {Pellat}},\ }\href@noop {}
  {\bibfield  {journal} {\bibinfo  {journal} {Astrophys. J.}\ }\textbf
  {\bibinfo {volume} {318}},\ \bibinfo {pages} {L43} (\bibinfo {year}
  {1987})}\BibitemShut {NoStop}%
\bibitem [{\citenamefont {Sygnet}\ \emph {et~al.}(1988)\citenamefont {Sygnet},
  \citenamefont {Tagger}, \citenamefont {Athanassoula},\ and\ \citenamefont
  {Pellat}}]{sygnet}%
  \BibitemOpen
  \bibfield  {author} {\bibinfo {author} {\bibfnamefont {J.~F.}\ \bibnamefont
  {Sygnet}}, \bibinfo {author} {\bibfnamefont {M.}~\bibnamefont {Tagger}},
  \bibinfo {author} {\bibfnamefont {E.}~\bibnamefont {Athanassoula}}, \ and\
  \bibinfo {author} {\bibfnamefont {R.}~\bibnamefont {Pellat}},\ }\href
  {\doibase 10.1093/mnras/232.4.733} {\bibfield  {journal} {\bibinfo  {journal}
  {Mon. Notices Royal Astron. Soc.}\ }\textbf {\bibinfo {volume} {232}},\
  \bibinfo {pages} {733} (\bibinfo {year} {1988})}\BibitemShut {NoStop}%
\bibitem [{\citenamefont {Masset}\ and\ \citenamefont
  {Tagger}(1999)}]{masset1999non}%
  \BibitemOpen
  \bibfield  {author} {\bibinfo {author} {\bibfnamefont {F.}~\bibnamefont
  {Masset}}\ and\ \bibinfo {author} {\bibfnamefont {M.}~\bibnamefont
  {Tagger}},\ }\href@noop {} {\bibfield  {journal} {\bibinfo  {journal} {arXiv
  preprint astro-ph/9902125}\ } (\bibinfo {year} {1999})}\BibitemShut {NoStop}%
\bibitem [{\citenamefont {Hamilton}\ \emph {et~al.}(2022)\citenamefont
  {Hamilton}, \citenamefont {Tolman}, \citenamefont {Arzamasskiy},\ and\
  \citenamefont {Duarte}}]{hamilton2022galactic}%
  \BibitemOpen
  \bibfield  {author} {\bibinfo {author} {\bibfnamefont {C.}~\bibnamefont
  {Hamilton}}, \bibinfo {author} {\bibfnamefont {E.~A.}\ \bibnamefont
  {Tolman}}, \bibinfo {author} {\bibfnamefont {L.}~\bibnamefont {Arzamasskiy}},
  \ and\ \bibinfo {author} {\bibfnamefont {V.~N.}\ \bibnamefont {Duarte}},\
  }\href@noop {} {\bibfield  {journal} {\bibinfo  {journal} {arXiv preprint
  arXiv:2208.03855}\ } (\bibinfo {year} {2022})}\BibitemShut {NoStop}%
\bibitem [{\citenamefont {Hamilton}(2023)}]{hamilton2023evolution}%
  \BibitemOpen
  \bibfield  {author} {\bibinfo {author} {\bibfnamefont {C.}~\bibnamefont
  {Hamilton}},\ }\href {\doibase 10.48550/arXiv.2302.06602} {\bibfield
  {journal} {\bibinfo  {journal} {arXiv preprint arXiv:2302.06602}\ } (\bibinfo
  {year} {2023}),\ 10.48550/arXiv.2302.06602}\BibitemShut {NoStop}%
\bibitem [{\citenamefont {Gorelenkov}\ \emph {et~al.}(2018)\citenamefont
  {Gorelenkov}, \citenamefont {Duarte}, \citenamefont {Podest{\`{a}}},\ and\
  \citenamefont {Berk}}]{GorelenkovDuarteNF2018}%
  \BibitemOpen
  \bibfield  {author} {\bibinfo {author} {\bibfnamefont {N.~N.}\ \bibnamefont
  {Gorelenkov}}, \bibinfo {author} {\bibfnamefont {V.~N.}\ \bibnamefont
  {Duarte}}, \bibinfo {author} {\bibfnamefont {M.}~\bibnamefont
  {Podest{\`{a}}}}, \ and\ \bibinfo {author} {\bibfnamefont {H.~L.}\
  \bibnamefont {Berk}},\ }\href
  {http://stacks.iop.org/0029-5515/58/i=8/a=082016} {\bibfield  {journal}
  {\bibinfo  {journal} {Nucl. Fusion}\ }\textbf {\bibinfo {volume} {58}},\
  \bibinfo {pages} {082016} (\bibinfo {year} {2018})}\BibitemShut {NoStop}%
\bibitem [{Note1()}]{Note1}%
  \BibitemOpen
  \bibinfo {note} {The equations presented in this paper are not exactly the
  same as the equations found in \cite {Zalesny}. Instead, they adhere to the
  original normalization of Berk \protect \textit {et al} \cite {BerkPRL1996}.
  The conversion between Berk's and Zalesny's normalization is $2 A_{\protect
  \text {Zalesny}} = A_{\protect \text {Berk}}$.}\BibitemShut {Stop}%
\bibitem [{\citenamefont {Duarte}\ and\ \citenamefont
  {Gorelenkov}(2019)}]{duarte2018analytical}%
  \BibitemOpen
  \bibfield  {author} {\bibinfo {author} {\bibfnamefont {V.~N.}\ \bibnamefont
  {Duarte}}\ and\ \bibinfo {author} {\bibfnamefont {N.~N.}\ \bibnamefont
  {Gorelenkov}},\ }\href {\doibase 10.1088/1741-4326/ab0135} {\bibfield
  {journal} {\bibinfo  {journal} {Nucl. Fusion}\ }\textbf {\bibinfo {volume}
  {59}},\ \bibinfo {pages} {044003} (\bibinfo {year} {2019})}\BibitemShut
  {NoStop}%
\bibitem [{\citenamefont {Duarte}\ \emph {et~al.}(2017)\citenamefont {Duarte},
  \citenamefont {Berk}, \citenamefont {Gorelenkov}, \citenamefont {Heidbrink},
  \citenamefont {Kramer}, \citenamefont {Nazikian}, \citenamefont {Pace},
  \citenamefont {Podest{\`{a}}}, \citenamefont {Tobias},\ and\ \citenamefont
  {Van~Zeeland}}]{DuarteAxivPRL}%
  \BibitemOpen
  \bibfield  {author} {\bibinfo {author} {\bibfnamefont {V.~N.}\ \bibnamefont
  {Duarte}}, \bibinfo {author} {\bibfnamefont {H.~L.}\ \bibnamefont {Berk}},
  \bibinfo {author} {\bibfnamefont {N.~N.}\ \bibnamefont {Gorelenkov}},
  \bibinfo {author} {\bibfnamefont {W.~W.}\ \bibnamefont {Heidbrink}}, \bibinfo
  {author} {\bibfnamefont {G.~J.}\ \bibnamefont {Kramer}}, \bibinfo {author}
  {\bibfnamefont {R.}~\bibnamefont {Nazikian}}, \bibinfo {author}
  {\bibfnamefont {D.~C.}\ \bibnamefont {Pace}}, \bibinfo {author}
  {\bibfnamefont {M.}~\bibnamefont {Podest{\`{a}}}}, \bibinfo {author}
  {\bibfnamefont {B.~J.}\ \bibnamefont {Tobias}}, \ and\ \bibinfo {author}
  {\bibfnamefont {M.~A.}\ \bibnamefont {Van~Zeeland}},\ }\href
  {http://stacks.iop.org/0029-5515/57/i=5/a=054001} {\bibfield  {journal}
  {\bibinfo  {journal} {Nucl. Fusion}\ }\textbf {\bibinfo {volume} {57}},\
  \bibinfo {pages} {054001} (\bibinfo {year} {2017})}\BibitemShut {NoStop}%
\bibitem [{\citenamefont {Duarte}\ \emph {et~al.}(2023)\citenamefont {Duarte},
  \citenamefont {Lestz}, \citenamefont {Gorelenkov},\ and\ \citenamefont
  {White}}]{duarte2020shifting}%
  \BibitemOpen
  \bibfield  {author} {\bibinfo {author} {\bibfnamefont {V.~N.}\ \bibnamefont
  {Duarte}}, \bibinfo {author} {\bibfnamefont {J.~B.}\ \bibnamefont {Lestz}},
  \bibinfo {author} {\bibfnamefont {N.~N.}\ \bibnamefont {Gorelenkov}}, \ and\
  \bibinfo {author} {\bibfnamefont {R.~B.}\ \bibnamefont {White}},\ }\href
  {\doibase 10.1103/PhysRevLett.130.105101} {\bibfield  {journal} {\bibinfo
  {journal} {Phys. Rev. Lett.}\ }\textbf {\bibinfo {volume} {130}},\ \bibinfo
  {pages} {105101} (\bibinfo {year} {2023})}\BibitemShut {NoStop}%
\bibitem [{\citenamefont {Barkley}\ \emph {et~al.}(2000)\citenamefont
  {Barkley}, \citenamefont {Tuckerman},\ and\ \citenamefont
  {Golubitsky}}]{Barkley}%
  \BibitemOpen
  \bibfield  {author} {\bibinfo {author} {\bibfnamefont {D.}~\bibnamefont
  {Barkley}}, \bibinfo {author} {\bibfnamefont {L.~S.}\ \bibnamefont
  {Tuckerman}}, \ and\ \bibinfo {author} {\bibfnamefont {M.}~\bibnamefont
  {Golubitsky}},\ }\href {\doibase 10.1103/PhysRevE.61.5247} {\bibfield
  {journal} {\bibinfo  {journal} {Phys. Rev. E}\ }\textbf {\bibinfo {volume}
  {61}},\ \bibinfo {pages} {5247} (\bibinfo {year} {2000})}\BibitemShut
  {NoStop}%
\bibitem [{\citenamefont {Sheard}\ \emph {et~al.}(2003)\citenamefont {Sheard},
  \citenamefont {Thompson},\ and\ \citenamefont {Hourigan}}]{Sheard}%
  \BibitemOpen
  \bibfield  {author} {\bibinfo {author} {\bibfnamefont {G.~J.}\ \bibnamefont
  {Sheard}}, \bibinfo {author} {\bibfnamefont {M.~C.}\ \bibnamefont
  {Thompson}}, \ and\ \bibinfo {author} {\bibfnamefont {K.}~\bibnamefont
  {Hourigan}},\ }\href {\doibase 10.1063/1.1597471} {\bibfield  {journal}
  {\bibinfo  {journal} {Phys. Fluids}\ }\textbf {\bibinfo {volume} {15}},\
  \bibinfo {pages} {L68} (\bibinfo {year} {2003})}\BibitemShut {NoStop}%
\bibitem [{\citenamefont {Lestz}\ and\ \citenamefont
  {Duarte}(2021)}]{LestzDuartePoP2021}%
  \BibitemOpen
  \bibfield  {author} {\bibinfo {author} {\bibfnamefont {J.~B.}\ \bibnamefont
  {Lestz}}\ and\ \bibinfo {author} {\bibfnamefont {V.~N.}\ \bibnamefont
  {Duarte}},\ }\href {\doibase 10.1063/5.0043979} {\bibfield  {journal}
  {\bibinfo  {journal} {Phys. Plasmas}\ }\textbf {\bibinfo {volume} {28}},\
  \bibinfo {pages} {062102} (\bibinfo {year} {2021})}\BibitemShut {NoStop}%
\bibitem [{Note2()}]{Note2}%
  \BibitemOpen
  \bibinfo {note} {A similar procedure can be done to obtain a second order
  differential equation in terms of the first wave amplitude. The indices for
  the second wave should then be switched with the first wave to arrive at this
  new equation.}\BibitemShut {Stop}%
\bibitem [{\citenamefont {Wiggins}(2003)}]{wiggins2003introduction}%
  \BibitemOpen
  \bibfield  {author} {\bibinfo {author} {\bibfnamefont {S.}~\bibnamefont
  {Wiggins}},\ }\href@noop {} {\emph {\bibinfo {title} {Introduction to applied
  nonlinear dynamical systems and chaos}}},\ Vol.~\bibinfo {volume} {2}\
  (\bibinfo  {publisher} {Springer},\ \bibinfo {year} {2003})\BibitemShut
  {NoStop}%
\bibitem [{\citenamefont {Lee}\ \emph {et~al.}(1998)\citenamefont {Lee},
  \citenamefont {Kwak},\ and\ \citenamefont {Lim}}]{LeeKwakLim1998}%
  \BibitemOpen
  \bibfield  {author} {\bibinfo {author} {\bibfnamefont {K.~J.}\ \bibnamefont
  {Lee}}, \bibinfo {author} {\bibfnamefont {Y.}~\bibnamefont {Kwak}}, \ and\
  \bibinfo {author} {\bibfnamefont {T.~K.}\ \bibnamefont {Lim}},\ }\href
  {\doibase 10.1103/PhysRevLett.81.321} {\bibfield  {journal} {\bibinfo
  {journal} {Phys. Rev. Lett.}\ }\textbf {\bibinfo {volume} {81}},\ \bibinfo
  {pages} {321} (\bibinfo {year} {1998})}\BibitemShut {NoStop}%
\bibitem [{\citenamefont {Heeter}\ \emph {et~al.}(2000)\citenamefont {Heeter},
  \citenamefont {Fasoli},\ and\ \citenamefont {Sharapov}}]{HeeterPRL2000}%
  \BibitemOpen
  \bibfield  {author} {\bibinfo {author} {\bibfnamefont {R.~F.}\ \bibnamefont
  {Heeter}}, \bibinfo {author} {\bibfnamefont {A.~F.}\ \bibnamefont {Fasoli}},
  \ and\ \bibinfo {author} {\bibfnamefont {S.~E.}\ \bibnamefont {Sharapov}},\
  }\href {\doibase 10.1103/PhysRevLett.85.3177} {\bibfield  {journal} {\bibinfo
   {journal} {Phys. Rev. Lett.}\ }\textbf {\bibinfo {volume} {85}},\ \bibinfo
  {pages} {3177} (\bibinfo {year} {2000})}\BibitemShut {NoStop}%
\bibitem [{\citenamefont {Bierwage}\ \emph {et~al.}(2017)\citenamefont
  {Bierwage}, \citenamefont {Shinohara}, \citenamefont {Todo}, \citenamefont
  {Aiba}, \citenamefont {Ishikawa}, \citenamefont {Matsunaga}, \citenamefont
  {Takechi},\ and\ \citenamefont {Yagi}}]{BierwageNF2017}%
  \BibitemOpen
  \bibfield  {author} {\bibinfo {author} {\bibfnamefont {A.}~\bibnamefont
  {Bierwage}}, \bibinfo {author} {\bibfnamefont {K.}~\bibnamefont {Shinohara}},
  \bibinfo {author} {\bibfnamefont {Y.}~\bibnamefont {Todo}}, \bibinfo {author}
  {\bibfnamefont {N.}~\bibnamefont {Aiba}}, \bibinfo {author} {\bibfnamefont
  {M.}~\bibnamefont {Ishikawa}}, \bibinfo {author} {\bibfnamefont
  {G.}~\bibnamefont {Matsunaga}}, \bibinfo {author} {\bibfnamefont
  {M.}~\bibnamefont {Takechi}}, \ and\ \bibinfo {author} {\bibfnamefont
  {M.}~\bibnamefont {Yagi}},\ }\href
  {http://stacks.iop.org/0029-5515/57/i=1/a=016036} {\bibfield  {journal}
  {\bibinfo  {journal} {Nucl. Fusion}\ }\textbf {\bibinfo {volume} {57}},\
  \bibinfo {pages} {016036} (\bibinfo {year} {2017})}\BibitemShut {NoStop}%
\bibitem [{\citenamefont {Bierwage}\ \emph {et~al.}(2021)\citenamefont
  {Bierwage}, \citenamefont {White},\ and\ \citenamefont
  {Duarte}}]{bierwage2021effect}%
  \BibitemOpen
  \bibfield  {author} {\bibinfo {author} {\bibfnamefont {A.}~\bibnamefont
  {Bierwage}}, \bibinfo {author} {\bibfnamefont {R.~B.}\ \bibnamefont {White}},
  \ and\ \bibinfo {author} {\bibfnamefont {V.~N.}\ \bibnamefont {Duarte}},\
  }\href@noop {} {\bibfield  {journal} {\bibinfo  {journal} {Plasma Fusion
  Res.}\ }\textbf {\bibinfo {volume} {16}},\ \bibinfo {pages} {1403087}
  (\bibinfo {year} {2021})}\BibitemShut {NoStop}%
\bibitem [{\citenamefont {Kuramoto}(1975)}]{kuramoto1975self}%
  \BibitemOpen
  \bibfield  {author} {\bibinfo {author} {\bibfnamefont {Y.}~\bibnamefont
  {Kuramoto}},\ }in\ \href@noop {} {\emph {\bibinfo {booktitle} {International
  Symposium on Mathematical Problems in Theoretical Physics: January 23--29,
  1975, Kyoto University, Kyoto/Japan}}}\ (\bibinfo {organization} {Springer},\
  \bibinfo {year} {1975})\ pp.\ \bibinfo {pages} {420--422}\BibitemShut
  {NoStop}%
\bibitem [{\citenamefont {Acebr{\'o}n}\ \emph {et~al.}(2005)\citenamefont
  {Acebr{\'o}n}, \citenamefont {Bonilla}, \citenamefont {Vicente},
  \citenamefont {Ritort},\ and\ \citenamefont {Spigler}}]{acebron2005kuramoto}%
  \BibitemOpen
  \bibfield  {author} {\bibinfo {author} {\bibfnamefont {J.~A.}\ \bibnamefont
  {Acebr{\'o}n}}, \bibinfo {author} {\bibfnamefont {L.~L.}\ \bibnamefont
  {Bonilla}}, \bibinfo {author} {\bibfnamefont {C.~J.~P.}\ \bibnamefont
  {Vicente}}, \bibinfo {author} {\bibfnamefont {F.}~\bibnamefont {Ritort}}, \
  and\ \bibinfo {author} {\bibfnamefont {R.}~\bibnamefont {Spigler}},\
  }\href@noop {} {\bibfield  {journal} {\bibinfo  {journal} {Rev. Mod. Phys.}\
  }\textbf {\bibinfo {volume} {77}},\ \bibinfo {pages} {137} (\bibinfo {year}
  {2005})}\BibitemShut {NoStop}%
\bibitem [{\citenamefont {Heeter}(1999)}]{HeeterThesis}%
  \BibitemOpen
  \bibfield  {author} {\bibinfo {author} {\bibfnamefont {R.~F.}\ \bibnamefont
  {Heeter}},\ }\href@noop {} {\bibfield  {journal} {\bibinfo  {journal}
  {\textit{"Alfven Eigenmode and Ion Bernstein Wave Studies for Controlling
  Fusion Alpha Particles"}, PhD thesis, Princeton University}\ } (\bibinfo
  {year} {1999})}\BibitemShut {NoStop}%
\bibitem [{\citenamefont {Liu}\ \emph {et~al.}(2022)\citenamefont {Liu},
  \citenamefont {Wei}, \citenamefont {Lin}, \citenamefont {Brochard},
  \citenamefont {Choi}, \citenamefont {Heidbrink}, \citenamefont {Nicolau},\
  and\ \citenamefont {McKee}}]{LiuPRL2022}%
  \BibitemOpen
  \bibfield  {author} {\bibinfo {author} {\bibfnamefont {P.}~\bibnamefont
  {Liu}}, \bibinfo {author} {\bibfnamefont {X.}~\bibnamefont {Wei}}, \bibinfo
  {author} {\bibfnamefont {Z.}~\bibnamefont {Lin}}, \bibinfo {author}
  {\bibfnamefont {G.}~\bibnamefont {Brochard}}, \bibinfo {author}
  {\bibfnamefont {G.~J.}\ \bibnamefont {Choi}}, \bibinfo {author}
  {\bibfnamefont {W.~W.}\ \bibnamefont {Heidbrink}}, \bibinfo {author}
  {\bibfnamefont {J.~H.}\ \bibnamefont {Nicolau}}, \ and\ \bibinfo {author}
  {\bibfnamefont {G.~R.}\ \bibnamefont {McKee}},\ }\href {\doibase
  10.1103/PhysRevLett.128.185001} {\bibfield  {journal} {\bibinfo  {journal}
  {Phys. Rev. Lett.}\ }\textbf {\bibinfo {volume} {128}},\ \bibinfo {pages}
  {185001} (\bibinfo {year} {2022})}\BibitemShut {NoStop}%
\bibitem [{Note3()}]{Note3}%
  \BibitemOpen
  \bibinfo {note} {The general procedure to solving an equation of Li\'{e}nard
  form is to use a variable transformation to reduce to an Abel's equation of
  the second kind. Then, another variable transform reduces the equation to a
  Bernoulli differential equation. For details of such procedure, refer to A.
  D. Polyanin and V. F. Zaitsev, \protect \textit {Handbook of Nonlinear
  Partial Differential Equations: Exact Solutions, Methods, and Problems}.
  Chapman and Hall/CRC, 2003.}\BibitemShut {Stop}%
\end{thebibliography}%
\bibliographystyle{apsrev4-1}

\begin{widetext}
\section*{Appendix}

\subsection{Calculation of Reduced Coefficients}
In order to calculate the values of the reduced coefficients, the integral is sectioned into simpler integrals. Eq. \ref{amp1} is reproduced below. A summation over the constants derived from parts from Sections A, B, and C results in the constants displayed in Eq. \ref{constants}. In this case, $b_1 = b_{1,0} + b_{1,1} + b_{1,2}$.

\begin{equation*}
\begin{aligned}
\frac{\mathrm{d} \hat{A_1}}{\mathrm{d} t}= & \hat{A_1} - \frac{1}{2} \int_0^{t/2} \mathrm{d}\eta\int_0^{t - 2\eta}\mathrm{d}\chi \cdot \\
& \bigl[\underbrace{\eta^2 \cdot \bigl(\hat{A_1}(t - \eta)\hat{A_1}(t - \eta - \chi)\hat{A_1^*}(t - 2\eta - \chi)}_{\textbf{Integral A}}\\
& + \underbrace{\hat{A_1}(t- \eta)\hat{A_2}(t - \eta - \chi)\hat{A_2^*}(t - 2 \eta - \chi) \cdot e^{-ip_1\eta}}_{\textbf{Integral B} }\bigr)\\
& + \underbrace{\hat{A_2}(t - \eta)\hat{A_1}(t - \eta - \chi)\hat{A_2^*}(t - 2\eta - \chi) e^{-ip_1(2\eta + \chi)} \cdot \eta (\eta + u_1(\eta + \chi ))}_{\textbf{Integral C}}\bigr] \cdot e^{-\hat{\nu} (2\eta + \chi)}
\end{aligned}
\end{equation*}
\subsubsection{Integral A}
\begin{equation}
\begin{aligned}
    & \frac{1}{2}\int_0^{t/2}\mathrm{d} \eta \int_0^{t-2\eta}\mathrm{d}\chi \ \eta^2\hat{A_1}(t - \eta)\hat{A_1}(t - \eta - \chi)\hat{A_1^*}(t - 2\eta - \chi) e^{-\hat{\nu}(2\eta + \chi)}  \\
    \approx \ & -\frac{1}{2\hat\nu}\hat{A_1}(t)|\hat{A_1}(t)|^2\int_0^{t/2}\mathrm{d} \eta \eta^2\left(e^{-\hat{\nu} t } - e^{-2\hat{\nu} \eta}\right) \\
    \approx \ & \frac{1}{2\hat\nu^4} \hat{A_1}(t)|\hat{A_1}(t)|^2\int_0^\infty x^2e^{-2x} \mathrm{d}x \ ; \ x = \hat{\nu} \eta \\
    = \ & b_0 \hat{A_1}(t)|\hat{A_1}(t)|^2 \implies b_0 = \left(8\hat\nu^4\right)^{-1}
\end{aligned}
\end{equation}

\subsubsection{Integral B}

\begin{equation}
\begin{aligned}
     & \frac{1}{2}\int_0^{t/2}\mathrm{d} \eta \int_0^{t-2\eta}\mathrm{d}\chi \ \eta^2\hat{A_1}(t- \eta)\hat{A_2}(t - \eta - \chi)\hat{A_2^*}(t - 2 \eta - \chi) e^{-ip_1\eta}e^{-\hat{\nu}(2\eta + \chi)} \\
     \approx \ & -\frac{1}{2\hat\nu}\int_0^{t/2}\mathrm{d} \eta \eta^2\hat{A_1}(t)|\hat{A_2}(t)|^2 e^{-ip_1\eta} \left(e^{-\hat{\nu}t} - e^{-\hat{\nu}(2\eta)}\right) \\
     \approx \ & \frac{1}{2\hat\nu^4}\hat{A_1}(t)|\hat{A_2}(t)|^2\int_0^\infty x^2 e^{-\xi x}\mathrm{d} x \ \ \left(x = \hat{\nu}\eta \ ; \xi = 2 + \frac{p_1}{\hat{\nu}}i \right) \\
     = \ & b_{1,0} \hat{A_1}(t)|\hat{A_2}(t)|^2 \implies b_{1,0} = \frac{1}{8\hat\nu^4\xi^3} = \frac{1}{8\hat\nu^4}\left(2+ \frac{p_1}{\hat{\nu}}i\right)^{-3}
\end{aligned}
\end{equation}

\subsubsection{Integral C}


\begin{equation}
    \begin{aligned}
         & \frac{1}{2} \int_0^{t/2} \mathrm{d}\eta\int_0^{t - 2\eta}\mathrm{d}\chi \hat{A_2}(t - \eta)\hat{A_1}(t - \eta - \chi)\hat{A_2^*}(t - 2\eta - \chi) \cdot \eta^2(1+u_1)\cdot e^{-(\hat{\nu} + p_1 i) (2\eta + \chi)}  \\
        \approx \ & -\frac{1}{2} \int_0^{t/2} \mathrm{d}\eta\hat{A_1}(t)|\hat{A_2}(t)|^2 \cdot \eta^2(1+u_1)(\hat{\nu} + p_1 i)^{-1}\cdot \left(e^{-(\hat{\nu} + p_1 i)t} - e^{-(\hat{\nu} + p_1 i)(2\eta)}\right) \\
        \approx \ & \frac{1}{2\hat\nu^4\Gamma} \ \hat{A_1}(t)|\hat{A_2}(t)|^2 \cdot (1+u_1) \int_0^\infty x^2e^{-2\Gamma x}\mathrm{d} x \ \ \left(x = \hat{\nu}\eta \ ; \Gamma = 1 + \frac{p_1}{\hat{\nu}}i \right) \\
        = \ & b_{1,1}\hat{A_1}(t)|\hat{A_2}(t)|^2 \implies b_{1,1}  = \frac{1+u_1}{8\hat{\nu}^{4}\Gamma^4}= \frac{1+u_1}{8\hat\nu^4}\left(1 + \frac{p_1}{\hat{\nu}} i\right)^{-4}
    \end{aligned}
\end{equation}

\begin{equation}
    \begin{aligned}
         & \frac{1}{2} \int_0^{t/2} \mathrm{d}\eta\int_0^{t - 2\eta}\mathrm{d}\chi \hat{A_2}(t - \eta)\hat{A_1}(t - \eta - \chi)\hat{A_2^*}(t - 2\eta - \chi) \cdot \eta u_1 \chi \cdot e^{-(\hat{\nu} + p_1 i) (2\eta + \chi)}   \\
        \approx \ & -\frac{1}{2} \int_0^{t/2} \mathrm{d}\eta\hat{A_1}(t)|\hat{A_2}(t)|^2 u_1 \eta \cdot (\hat{\nu} + p_1 i)^{-1}\left(\left.\chi e^{-(\hat{\nu} + p_1 i)(2\eta + \chi)}\right|_0^{t-2\eta} - \int_0^{t-2\eta}\mathrm{d}\chi e^{-(\hat{\nu} + p_1 i)(2\eta + \chi)}\right) \\
        = \ & -\frac{1}{2} \int_0^{t/2} \mathrm{d}\eta\hat{A_1}(t)|\hat{A_2}(t)|^2 u_1 \eta \cdot (\hat{\nu} + p_1 i)^{-1}\left[ (t- 2\eta) e^{-(\hat{\nu} + p_1 i)t} + (\hat{\nu} + p_1 i)^{-1} \left(e^{-(\hat{\nu} + p_1 i)t} - e^{-(\hat{\nu} + p_1 i)(2\eta)}\right)\right]\\
        \approx \ & \frac{1}{2\hat\nu^4\Gamma^2} \ \hat{A_1}(t)|\hat{A_2}(t)|^2 \cdot u_1 \int_0^\infty xe^{-2\Gamma x}\mathrm{d} x \ \ \left(x = \hat{\nu}\eta \ ; \Gamma = 1 + \frac{p_1}{\hat{\nu}}i \right) \\
        = \ & b_{1,2}\hat{A_1}(t)|\hat{A_2}(t)|^2 \implies b_{1,2} = \frac{u_1}{8\hat{\nu}^{4}\Gamma^4} = \frac{u_1}{8\hat\nu^4}\left(1 + \frac{p_1}{\hat{\nu}}i\right)^{-4}
    \end{aligned}
\end{equation}

\subsection{Analytic Solutions of Eq. \ref{eq:secondorderform} for Special Cases }

\subsubsection{Reduction to Li\'{e}nard Form}

Eq. \ref{eq:secondorderform} has the form of a Li\'{e}nard equation, which can be analytically solvable in some special cases. The general form of Li\'{e}nard's equation is $y'' = [a(2n+k)y^k + b]y^{n-1}y' +(-a^2ny^{2k}-aby^k + c)y^{2n-1}$, where $a$, $b$, $c$, $k$, and $n$ are constants. 
To analytically solve the differential equation of Li\'{e}nard form, two variable transformations are required to simplify. \footnote{The general procedure to solving an equation of Li\'{e}nard form is to use a variable transformation to reduce to an Abel's equation of the second kind. Then, another variable transform reduces the equation to a Bernoulli differential equation. For details of such procedure, refer to A. D. Polyanin and V. F. Zaitsev, \textit{Handbook of Nonlinear Partial Differential Equations: Exact Solutions, Methods, and Problems}. Chapman and Hall/CRC, 2003.}

Eq. \ref{eq:secondorderform} can be recast into the following form: $\Psi'' = (h_1\Psi^k + h_2)\Psi^{n-1}\Psi' + (h_3\Psi^{2k} + h_4\Psi^k + h_5)\Psi^{2n-1}$ where $h_{j}$, $k$, and $n$ are constants. For matching purposes, $k=-\mathrm{Re}(b_2)/b_0$ and $n=1$. The constants $h_j$ are also matched in Eq.~\ref{eq:hvals}.

\begin{equation} \label{eq:hvals}
    \begin{aligned}
        & h_1 = -2\left(b_0 + \mathrm{Re}(b_1) -\frac{2b_0^2}{\mathrm{Re}(b_2)}\right) = a\left(2 -\frac{\mathrm{Re}(b_2)}{b_0}\right) \\
        & h_2 = 2\left(1-\frac{2b_0}{\mathrm{Re}(b_2)}\right) = b \\
        & h_3 = -\frac{4b_0^2}{\mathrm{Re}(b_2)}\left(\frac{b_0^2}{\mathrm{Re}(b_2)} - \mathrm{Re}(b_1)\right) = -a^2 \\
        & h_4 = \frac{4b_0}{\mathrm{Re}(b_2)} \left(\frac{2b_0^2}{\mathrm{Re}(b_2)} - \mathrm{Re}(b_1) - b_0\right) = -ab \\
        & h_5 = \frac{4b_0}{\mathrm{Re}(b_2)} \left(1 - \frac{b_0}{\mathrm{Re}(b_2)}\right) = c \\
    \end{aligned}
\end{equation}

To solve this system of equations for $a$, $b$, and $c$, there requires at maximum two equations of constraint for the variables $b_0$, $\mathrm{Re}(b_1)$, and $\mathrm{Re}(b_2)$. In this case, there is only one constraint which is $\mathrm{Re}(b_1)-2b_0+\mathrm{Re}(b_2)=0$. If this criteria is satisfied, there exists an analytical form for the time evolution of the wave amplitude. The constants calculated in Li\'{e}nard equation are presented in Eq. \ref{eq:Lienardcons1}.

\begin{equation}
    \begin{aligned}
       & a = -\frac{2}{2 -\frac{\mathrm{Re}(b_2)}{b_0}}\left(b_0 + \mathrm{Re}(b_1) -\frac{2b_0^2}{\mathrm{Re}(b_2)}\right)\\
       & b = 2\left(1-\frac{2b_0}{\mathrm{Re}(b_2)}\right)\\
       &c = \frac{4b_0}{\mathrm{Re}(b_2)} \left(1 - \frac{b_0}{\mathrm{Re}(b_2)}\right)
    \end{aligned}
    \label{eq:Lienardcons1}
\end{equation} 

\subsubsection{Alternate Form of Li\'{e}nard Equation}
The same treatment is applied to the alternate equation of Li\'{e}nard form to find an analytical solution.  The alternate form of Li\'{e}nard equation is $y'' = [a(2m + k)y^{2k} + b(2m-k)]y^{m-k-1}y'-(a^2my^{4k} + cy^{2k} + b^2m)y^{2m-2k-1}$ where $a$, $b$, $c$, $m$, and $k$ are constants.

Once again, Eq.~\ref{eq:secondorderform} can be recast in the following form $\Psi'' = (g_1\Psi^{2k} + g_2)\Psi^{m-k-1}\Psi' + (g_3\Psi^{4k} + g_4\Psi^{2k} + g_5)\Psi^{2m-2k-1}$. To match the recasted form to the general form of the alternate Li\'{e}nard equation, $k=-\mathrm{Re}(b_2)/2a_0$ and $m=1-\mathrm{Re}(b_2)/2a_0$.
\begin{equation}
    \begin{aligned}
        &g_1 = -2\left(b_0 + \mathrm{Re}(b_1) -\frac{2b_0^2}{\mathrm{Re}(b_2)}\right) = a\left[2\left(1-\frac{\mathrm{Re}(b_2)}{2b_0}\right)-\frac{\mathrm{Re}(b_2)}{2b_0}\right] =  a\left(2-\frac{3\mathrm{Re}(b_2)}{2b_0}\right)\\
        &g_2 = 2\left(1-\frac{2b_0}{\mathrm{Re}(b_2)}\right) = b\left[2\left(1-\frac{\mathrm{Re}(b_2)}{2b_0}\right)+\frac{\mathrm{Re}(b_2)}{2b_0}\right] = b\left(2-\frac{\mathrm{Re}(b_2)}{2b_0}\right)\\
        &g_3 = -\frac{4b_0^2}{\mathrm{Re}(b_2)}\left(\frac{b_0^2}{\mathrm{Re}(b_2)} - \mathrm{Re}(b_1)\right) = -a^2\left(1-\frac{\mathrm{Re}(b_2)}{2b_0}\right)\\
        &g_4 = \frac{4b_0}{\mathrm{Re}(b_2)} \left(\frac{2b_0^2}{\mathrm{Re}(b_2)} - \mathrm{Re}(b_1) - b_0\right) = -c\\
        &g_5 = \frac{4b_0}{\mathrm{Re}(b_2)} \left(1 - \frac{b_0}{\mathrm{Re}(b_2)}\right) = -b^2\left(1-\frac{\mathrm{Re}(b_2)}{2b_0}\right)\\
    \end{aligned}
\end{equation}

For this alternate form, there are two restraints on the variables $b_0$, $\mathrm{Re}(b_1)$, and $\mathrm{Re}(b_2)$. In this form, the alternate Li\'{e}nard's equation can be solved for select values: $\{\mathrm{Re}(b_1)/b_0,\mathrm{Re}(b_2)/b_0\} = \{ (\frac{1}{2},3), (3,3)\}$. The corresponding constants are presented in Table \ref{table:altLieconstants}. 

\begin{table}[th!]
\renewcommand{\arraystretch}{1.25}
\caption{Evaluation for Constants of Li\'{e}nard's Alternate Form \label{table:altLieconstants}}
\begin{ruledtabular}
\begin{tabular}{|c||ccc|}
$\left(\frac{\mathrm{Re}(b_1)}{b_0},\frac{\mathrm{Re}(b_2)}{b_0}\right)$ & $a$ & $b$ & $c$ \\
\hline \hline 
$\left(\frac{1}{2}, 3\right)$ & $\frac{2}{3}b_0$ & $\frac{4}{3}$ & $\frac{10}{9}b_0$ \\ [0.15cm] 
$\left(3,3\right)$ & $\frac{8}{3}b_0$ & $\frac{4}{3}$ & $\frac{40}{9}b_0$ \\
\end{tabular}
\end{ruledtabular}
\end{table}

 For completeness, the expressions for calculating the constants for the alternate form and the constraint equations are presented in Eq.~\ref{eq:altLconst} and \ref{eq:altLres} respectively.

\begin{equation} \label{eq:altLconst}
    \begin{aligned}
        &a =-\frac{2}{2-\frac{3\mathrm{Re}(b_2)}{2b_0}}\left(b_0 + \mathrm{Re}(b_1) -\frac{2b_0^2}{\mathrm{Re}(b_2)}\right) \\
        &b = \frac{2}{2-\frac{\mathrm{Re}(b_2)}{2b_0}}\left(1-\frac{2b_0}{\mathrm{Re}(b_2)}\right) \\
        &c = -\frac{4b_0}{\mathrm{Re}(b_2)} \left(\frac{2b_0^2}{\mathrm{Re}(b_2)} - \mathrm{Re}(b_1) - b_0\right)
    \end{aligned}
\end{equation}

\begin{equation} \label{eq:altLres}
    \begin{aligned}
         &b_0\left(4b_0-3\mathrm{Re}(b_2)\right)^2\left(b_0^2 - \mathrm{Re}(b_1)\mathrm{Re}(b_2)\right) = 2\bigl(b_0\mathrm{Re}(b_2) + \mathrm{Re}(b_1)\mathrm{Re}(b_2) -2b_0^2\bigr)^2(2b_0-\mathrm{Re}(b_2)) \\
        & (4b_0-\mathrm{Re}(b_2))^2 (\mathrm{Re}(b_2)-b_0) = -2\bigl[\mathrm{Re}(b_2)-2b_0\bigr]^2(2b_0-\mathrm{Re}(b_2))
    \end{aligned}
\end{equation}

\subsection{Numerical Methodology}
Simulating a time-delayed nonlinear integrodifferential equation can be computationally intensive in the general case. To help reduce computation time, a reduced recursive algorithm is constructed for two waves as a generalization of Heeter's scheme for a single wave \cite{HeeterThesis}. It takes advantage of the structure of the time delays to avoid recomputing previous terms of a sum at later times. The original cubic equation is reproduced below for the wave amplitude equation of the first wave. A similar procedure can be done for the wave amplitude equation of the second wave.

\begin{equation*}
\begin{aligned}
\frac{\mathrm{d} \hat{A_1}}{\mathrm{d} t}= & \hat{A_1} - \frac{1}{2} \int_0^{t/2} \mathrm{d}\eta\int_0^{t - 2\eta}\mathrm{d}\chi \cdot  \bigl[\eta^2 \cdot \bigl(\hat{A_1}(t - \eta)\hat{A_1}(t - \eta - \chi)\hat{A_1^*}(t - 2\eta - \chi)\\
& + \hat{A_1}(t- \eta)\hat{A_2}(t - \eta - \chi)\hat{A_2^*}(t - 2 \eta - \chi) \cdot e^{-ip_1\eta}\bigr) \\
& + \hat{A_2}(t - \eta)\hat{A_1}(t - \eta - \chi)\hat{A_2^*}(t - 2\eta - \chi) e^{-ip_1(2\eta + \chi)} \cdot \eta (\eta + u_1(\eta + \chi ))\bigr] \cdot e^{-\hat{\nu} (2\eta + \chi)}
\end{aligned}
\end{equation*}

When the wave amplitude evolution equation is discretized in time, it can be numerically solved via an Euler's method. By applying the following substitutions $\{j = t, \ k = \eta, \ l = t- 2\eta - \chi\}$, the equation can be expressed in terms of four summations denoted by $S_{1-4}$. In turn, these summations can be simplified by representing them recursively in j. The functions $S_{1-4}$ are presented below in summation form and recursive form. The recursive form eliminates a summation in the calculations, thereby boosting processing speed.

\begin{equation*}
\begin{aligned}
    \Delta \hat{A_1}(j) = \hat{A_1}(j) \Delta t - \frac{1}{2} (\Delta t)^5 \displaystyle\sum_{k=1}^{j/2}k^2 \Bigl[& \hat{A_1}(j-k) \cdot S_1(j,k) + e^{-ip_1k\Delta t}\cdot \hat{A_1}(j-k) \cdot S_2(j,k) \\
    &+ (u_1+1) \cdot \hat{A_2}(j-k) \cdot S_3(j,k) + \frac{u_1}{k} \cdot \hat{A_2}(j-k) \cdot S_4(j,k)\Bigr]
\end{aligned}
\end{equation*}

\begin{equation*}
\begin{aligned}
    S_1(j,k) =& \displaystyle\sum_{l=0}^{j-2k} \exp[\hat{\nu}(l-j)\Delta t]\hat{A_1}(k+l)\hat{A_1^*}(l) \\
    S_2(j,k) =& \displaystyle\sum_{l=0}^{j-2k} \exp[\hat{\nu}(l-j)\Delta t]\hat{A_2}(k+l)\hat{A_2^*}(l) \\
    S_3(j,k) =& \displaystyle\sum_{l=0}^{j-2k} \exp[(\hat{\nu}+ip_1)(l-j)\Delta t]\hat{A_1}(k+l)\hat{A_2^*}(l) \\
    S_4(j,k) =& \displaystyle\sum_{l=0}^{j-2k} (j-2k-l)\exp[(\hat{\nu}+ip_1)(l-j)\Delta t]\hat{A_1}(k+l)\hat{A_2^*}(l) \\
\end{aligned}
\end{equation*}

\begin{equation*}
\begin{aligned}
    S_1(j,k) =& \ e^{-\hat{\nu}\Delta t}S_1(j-1,k) + e^{-2k\hat{\nu}\Delta t}A_1(j-k)A_1^*(j-2k)\\
    S_2(j,k) =& \  e^{-\hat{\nu}\Delta t}S_2(j-1,k) + e^{-2k\hat{\nu}\Delta t}A_2(j-k)A_2^*(j-2k)\\
    S_3(j,k) =& \  e^{-(\hat{\nu}+ip_1)\Delta t}S_3(j-1,k) + e^{-2k(\hat{\nu}+ip_1)\Delta t}\hat{A_1}(j-k)\hat{A_2^*}(j-2k) \\
    S_4(j,k) =& \  e^{-(\hat{\nu}+ip_1)\Delta t}\bigl[S_4(j-1,k) + S_3(j-1,k)\bigr]. \\
\end{aligned}
\end{equation*}

\end{widetext}
\end{document}